\documentclass[fleqn,usenatbib,useAMS]{Papers}
\usepackage{newtxtext,newtxmath}
\usepackage[T1]{fontenc}
\usepackage{subfloat}
\usepackage{makecell}
\usepackage[authoryear]{natbib}
\usepackage{float}
\usepackage{graphicx}	
\usepackage{amsmath}	

\title[Two-body collapse model (TBCM) and GSCH]{Two-body collapse model for self-gravitating flow of dark matter and generalized stable clustering hypothesis for pairwise velocity}

\author[Z. Xu]{
Zhijie (Jay) Xu,$^{1}$\thanks{E-mail: zhijie.xu@pnnl.gov; zhijiexu@hotmail.com}
\\
$^{1}$Physical and Computational Sciences Directorate, Pacific Northwest National Laboratory; Richland, WA 99352, USA\\
}

\date{Accepted XXX. Received YYY; in original form ZZZ}

\pubyear{2022}

\begin{document}
\label{firstpage}
\pagerange{\pageref{firstpage}--\pageref{lastpage}}
\maketitle

\begin{abstract}
Analytical tools are valuable to study gravitational collapse. However, solutions are hard to find due to the highly non-linear nature. Only a few simple but powerful tools exist so far. Two examples are the spherical collapse model (SCM) and stable clustering hypothesis (SCH). We present a new analytical tool based on the elementary step of inverse mass cascade in dark matter flow, i.e. a two-body collapse model (TBCM). TBCM plays the same role as harmonic oscillator in dynamics and can be fundamental to understand structure evolution. For convenience, TBCM is formulated for gravity with any potential exponent $n$ in a static background with a fixed damping ($n$=-1 for Newtonian gravity). The competition between gravity, expanding background (or damping), and angular momentum classifies two-body collapse into: 1) free fall collapse for weak angular momentum, where free fall time is greater if same system starts to collapse at earlier time; 2) equilibrium collapse for weak damping that persists longer in time, whose perturbative solutions lead to power-law evolution of system energy and momentum. Two critical values $\beta_{s1}=1$ and $\beta_{s2}=1/3\pi$ are identified that quantifies the competition between damping and gravity. Value $\beta_{s2}$ only exists for discrete values of potential exponent $n=(2-6m)/(1+3m)=$ -1,-10/7... for integer $m$. Critical density ratio ($\Delta_c=18\pi^2$) is obtained for $n$=-1 that is consistent with SCM. TBCM predicts angular velocity $\propto Hr^{-3/2}$ for two-body system of size $r$. The isothermal density is a result of infinitesimal halo lifetime or extremely fast mass accretion. TBCM is able to demonstrate SCH, i.e. mean pairwise velocity (first moment) $\langle\Delta u\rangle=-Hr$. A generalized SCH is developed for higher order moments $\langle\Delta u^{2m+1}\rangle=-(2m+1)\langle\Delta u^{2m}\rangle Hr$ that is validated by N-body simulation. Energy evolution in TBCM is independent of particle mass and energy equipartition does not apply. TBCM can be considered as a non-radial SCM. Both models predict the same critical density ratio, while TBCM contains much richer information.
\end{abstract}

\begin{keywords}
\vspace*{-15pt}
Dark matter; N-body simulations; Theoretical models
\end{keywords}

\begingroup
\let\clearpage\relax
\tableofcontents
\endgroup

\section{Introduction}
\label{sec:1}
Collisionless systems often show properties strongly suggesting common physical principles that control the system motion and evolution. The self-gravitating collisionless fluid dynamics (SG-CFD) is the study of motion of collisionless matter under its own gravity. The large-scale gravitational collapse of dark matter is an example of SG-CFD and the basis of standard models for the formation of large-scale structures. Structure formation starts from the gravitational collapse of small-scale density fluctuations and proceeds hierarchically in a "bottom-up" fashion with small structures merging into large structures. The same process can be described by a halo-mediated inverse mass cascade, where halos (building blocks) pass their mass onto larger and larger halos, until halo mass growth becomes dominant over the mass propagation \citep{Xu:2021-Inverse-mass-cascade-mass-function}. Halos are necessary to form for collisionless system with long-range interaction to maximize system entropy \citep{Xu:2021-The-maximum-entropy-distributi, Xu:2021-Mass-functions-of-dark-matter-}. The merging of halos is an elementary step in mass cascade and the focus of current paper. 

The hierarchical merging of structures is a fundamental and complex step for structure formation. In a finite time interval $\Delta t$, the hierarchical merging might involve multiple substructures merging into a single large structure. However, for an infinitesimal interval $dt$, that process should involve the merging of two and only two substructures \citep{Mo:2010-Galaxy-formation-and-evolution}. In this regard, the two-body gravitational collapse is an elementary and fundamental step for hierarchical structure formation during mass cascade. While the two-body problem in static background (no space expansion) without damping is well-known, a comprehensive understanding of the two-body collapse (TBCM) in expanding background seems not fully developed. In fact, the TBCM can be a powerful analytical tool to study the non-linear structure formation and provide many insights into the energy and momentum evolution of N-body system \citep{Xu:2022-The-evolution-of-energy--momen}. This is made possible with analytical solutions of TBCM by transforming the original two-body system in a comoving expanding background to an equivalent transformed system in a static background with a fixed damping. Results analytically obtained in the transformed system can be equivalently transformed back to the original comoving system. 

Despite the great success of large-scale \textit{N}-body simulations for structure formation, there are always motivations for finding analytical approaches to gravitational collapse. However, this can be extremely difficult due to the highly non-linear nature of structure formation. Nonetheless, a few simple but powerful analytical tools exist for structure evolution. The first example makes use a spherical symmetry of an over-density to formulate the gravitational collapse, i.e. a spherical collapse model (SCM). Developed by Gunn \& Gott \citep{Gunn:1972-Infall-of-Matter-into-Clusters} and Gunn \citep{Gunn:1977-Massive-Galactic-Halos--1--For} in 1970s, the first SCM model provides solutions for the collapse of a spherical mass shell surrounding an over-density with an uniform density. The self-similar spherical collapse model was later developed in 1980s to allow for a non-uniform initial density and collapse of new shells \citep{Fillmore:1984-Self-Similar-Gravitational-Col,Bertschinger:1985-Self-Similar-Secondary-Infall-}. The idea of SCM model was further developed to consider the effect of non-radial orbit by introducing an additional constant centrifugal force due to the non-radial motion \citep{White:1992-Models-for-Galaxy-Halos-in-an-,Nusser:2001-Self-similar-spherical-collaps}. The SCM predicts the critical density ratio of halos that has been widely used for the development of halo mass functions and density profiles \citep{Press:1974-Formation-of-Galaxies-and-Clus,Cooray:2002-Halo-models-of-large-scale-str}. Similar predictions were also extended to the ellipsoidal collapse \citep{Sheth:2001-Ellipsoidal-collapse-and-an-im,Sheth:2002-An-excursion-set-model-of-hier}. 

The second example assumes that on a sufficiently small scale, the clusters of mass particles are bound and stable with a fixed mean physical separation between particles, i.e. a stable clustering hypothesis (SCH) \citep{Peebles:1974-The-gravitational-instability-,Davis:1977-Integration-of-Bbgky-Equations}. There is no stream motion between particles in physical coordinate. In this sense, the peculiar motion cancels out the Hubble flow and the hypothesis equivalently states that the mean (first order moment) pairwise peculiar velocity is proportional to the separation \textit{r} (physical distance) as $\left\langle \Delta u_{L} \right\rangle =-Hr$. The stable clustering hypothesis is a fundamental assumption for the nonlinear gravitational collapse at small scales. Combined with pair conservation equation \citep{Peebles:1980-The-Large-Scale-Structure-of-t}, the hypothesis can be used to predict the dynamic evolution of density correlation function on small scales. While directly proving SCH based on fundamental rules seems challenging, there have been many attempts to verify this assumption with \textit{N}-body simulations \citep{Efstathiou:1988-Gravitational-Clustering-from-,Colombi:1996-Self-similarity-and-scaling-be}. The limited resolution of simulations makes it difficult to achieve a sufficiently high accuracy at small sales where this assumption is valid. This paper provides a proof of original stable clustering hypothesis (SCH) based on the analytical solution of two-body collapse model (TBCM) and extends SCH to high order moments of pairwise velocity.  

The mass and energy cascade \citep{Xu:2021-Inverse-mass-cascade-mass-function,Xu:2021-Inverse-and-direct-cascade-of-} involve a series of elemetary two-body collapse, i.e. a chain reaction description \citep[see][Fig. 8]{Xu:2021-Inverse-mass-cascade-mass-function}. Understanding the cascade process is critical for the development of halo energy and momentum evolution \citep{Xu:2022-The-mean-flow--velocity-disper,Xu:2022-The-evolution-of-energy--momen} and the statistical theory for dark matter flow \citep{Xu:2022-The-statistical-theory-of-2nd,Xu:2022-The-statistical-theory-of-3rd,Xu:2022-Two-thirds-law-for-pairwise-ve}. In addition, the two-body collapse based mass and energy cascade are also potentially relevant to the dark matter particle mass and properties \citep{Xu:2022-Postulating-dark-matter-partic}, MOND (modified Newtonian dynamics) theory \citep{Xu:2022-The-origin-of-MOND-acceleratio}, and baryonic-to-halo mass relation \citep{Xu:2022-The-baryonic-to-halo-mass-rela}.  

In this paper, the elementary step of mass cascade (two-body collapse model -- TBCM) is mathematically formulated to provide another useful analytical tool and more insights into the structure formation and evolution. The TBCM model can demonstrate the standard stable clustering hypothesis on small scale for the first moment of pairwise velocity. A generalized stable clustering hypothesis (GSCH) can be subsequently derived for high order moments of pairwise velocity. The connections of TBCM with other analytical tools, including violent relaxation and spherical collapse model (SCM), are also discussed in detail. Both leads to the same prediction of critical halo density ratio, while TBCM contains much richer information. 

The rest of the paper is organized as follows: Section \ref{sec:2} introduces the equations of motion for the dynamics of a \textit{N}-body system. Equivalence is established between the original comoving system in expanding background and a transformed system in static background. The elementary gravitational collapse (TBCM model) is formulated and analytically solved in Section \ref{sec:3}, along with the applications of TBCM to identify distinct regimes and critical values. Connections with stable clustering hypothesis, violent relaxation and spherical collapse models are discussed in Section \ref{sec:4}.

\section{Equations of motion in comoving and transformed systems}
\label{sec:2}
In this section, the equivalence is first established between a comoving system in expanding background and a transformed system in static background. The self-gravitating of a system of \textit{N} collisionless particles in expanding background can be studied by solving governing equation of motion \citep[see][p. 44]{Peebles:1980-The-Large-Scale-Structure-of-t} in a comoving system (comoving coordinates $\boldsymbol{\mathrm{x}}$ and physical time \textit{t}) as
\begin{equation} 
\label{eq:1} 
\frac{d^{2} \boldsymbol{\mathrm{x}}_{i} }{dt^{2} } +2H\frac{d\boldsymbol{\mathrm{x}}_{i} }{dt} =-\frac{Gm_{p} }{a^{3} } \sum _{j\ne i}^{N}\frac{\boldsymbol{\mathrm{x}}_{i} -\boldsymbol{\mathrm{x}}_{j} }{\left|\boldsymbol{\mathrm{x}}_{i} -\boldsymbol{\mathrm{x}}_{j} \right|^{3} }  ,    
\end{equation} 
where $\boldsymbol{\mathrm{x}}_{i} $ is the comoving coordinate of \textit{N} particles with equal mass $m_{p} $ and $G$ is the standard gravitational constant. The Hubble constant $H\left(t\right)={\dot{a}/a} $, where \textit{a} is the scale factor. 

For growing halos from continuous mass accretion, an effective gravitational potential exponent $n_{e} \approx -1.3$ can be different from -1 for standard gravitational potential due to the finite halo surface energy \citep[see][Eq. (96)]{Xu:2021-Inverse-mass-cascade-halo-density}. This hints that it might be beneficial by looking at a general potential with an arbitrary exponent \textit{n}. The maximum entropy distributions of velocity and energy in SG-CFD have been developed for the long-range power-law potential with any exponent \textit{n} in (Xu 2021c). In this paper, we assume the same power-law gravitational potential $V_{p} $ with an arbitrary exponent of \textit{n }for particle-particle interacting, i.e. $V_{p} \left(r\right){=-G_{n} m_{p}^{2} /r^{-n} } $. Here $G_{n} $ is a generalized gravitational constant ($G_{n} =G$ when $n=-1$). The equation of motion with arbitrary exponent \textit{n} reads
\begin{equation} 
\label{ZEqnNum157264} 
\frac{d^{2} \boldsymbol{\mathrm{x}}_{i} }{dt^{2} } +2H\frac{d\boldsymbol{\mathrm{x}}_{i} }{dt} =\frac{nG_{n} m_{p} }{a^{3} } \sum _{j\ne i}^{N}\frac{\boldsymbol{\mathrm{x}}_{i} -\boldsymbol{\mathrm{x}}_{j} }{\left|\boldsymbol{\mathrm{x}}_{i} -\boldsymbol{\mathrm{x}}_{j} \right|^{2-n} }  .  
\end{equation} 
Let's introduce a new transformed time scale \textit{s} as ${ds/dt} =a^{p} $, where \textit{p} is an arbitrary exponent. The original Eq. \eqref{ZEqnNum157264} can be equivalently transformed to
\begin{equation} 
\label{ZEqnNum168751} 
\frac{d^{2} \boldsymbol{\mathrm{x}}_{i} }{ds^{2} } +\frac{d\boldsymbol{\mathrm{x}}_{i} }{ds} \left(p+2\right)a^{-p} H=\frac{nG_{n} m_{p} }{a^{3+2p} } \sum _{j\ne i}^{N}\frac{\boldsymbol{\mathrm{x}}_{i} -\boldsymbol{\mathrm{x}}_{j} }{\left|\boldsymbol{\mathrm{x}}_{i} -\boldsymbol{\mathrm{x}}_{j} \right|^{2-n} }  .      
\end{equation} 
Obviously $s=t$ if $p=0$ and Eq. \eqref{ZEqnNum168751} is reduced to Eq. \eqref{ZEqnNum157264}. Specifically, $p=-2$ eliminates the first order derivative and \textit{s} is the time variable for integration of \textit{N}-body simulation that allows for a symplectic (phase space volume preserving) integrator. Time scale \textit{s} becomes conformal time if $p=-1$. Another special case can be identified with $p=-{3/2} $ for a matter-dominant model,
\begin{equation}
H_{0}^{2} =H^{2} a^{3}, \quad \frac{dH}{dt} =-\frac{3}{2} H^{2}, \quad \textrm{and} \quad H^{2} =\frac{8\pi G\bar{\rho }_{y} \left(a\right)}{3},
\label{ZEqnNum377712}
\end{equation}
\noindent where $H_{0} $ is the Hubble constant at the present epoch (\textit{a}=1) and $\bar{\rho }_{y} \left(a\right)$ is the physical density of the homogeneous background. 

For $p=-{3/2} $ with Eq. \eqref{ZEqnNum377712}, Eq. \eqref{ZEqnNum168751} now becomes 
\begin{equation} 
\label{ZEqnNum730753} 
\frac{d^{2} \boldsymbol{\mathrm{x}}_{i} }{ds^{2} } +\frac{1}{2} H_{0} \frac{d\boldsymbol{\mathrm{x}}_{i} }{ds} =nG_{n} m_{p} \sum _{j\ne i}^{N}\frac{\boldsymbol{\mathrm{x}}_{i} -\boldsymbol{\mathrm{x}}_{j} }{\left|\boldsymbol{\mathrm{x}}_{i} -\boldsymbol{\mathrm{x}}_{j} \right|^{2-n} }  =\frac{\boldsymbol{\mathrm{F}}_{i} }{m_{p} } ,      
\end{equation} 
where $\boldsymbol{\mathrm{F}}_{i} $ is the resultant force on particle \textit{i} in comoving system. Clearly, the scale factor \textit{a} does not explicitly appear in Eq. \eqref{ZEqnNum730753} and the Hubble constant $H_{0} $ can be considered as a constant damping that is time-invariant. The original Eq. \eqref{ZEqnNum157264} in expanding background is now equivalently converted to a transformed system in static background with a constant damping ${H_{0} /2} $ (Eq. \eqref{ZEqnNum730753}) evolving with a new time scale \textit{s}. The transformed system consists of a comoving spatial coordinate $\boldsymbol{\mathrm{x}}_{i}$ and a transformed time scale \textit{s}. 

The particle velocity $\boldsymbol{\mathrm{v}}_{i}$ for transformed system can be written as, 
\begin{equation} 
\label{ZEqnNum631619} 
\boldsymbol{\mathrm{v}}_{i} =\frac{d\boldsymbol{\mathrm{x}}_{i} }{ds} =a^{{3/2} } \frac{d\boldsymbol{\mathrm{x}}_{i} }{dt} =a^{{1/2} } \boldsymbol{\mathrm{u}}_{i} ,          
\end{equation} 
while the peculiar velocity $\boldsymbol{\mathrm{u}}_{i} $ in physical time \textit{t} can be related to the new velocity $\boldsymbol{\mathrm{v}}_{i} $,
\begin{equation} 
\label{eq:7} 
\boldsymbol{\mathrm{u}}_{i} =a\frac{d\boldsymbol{\mathrm{x}}_{i} }{dt} =\frac{d\boldsymbol{\mathrm{r}}_{i} }{dt} -H\boldsymbol{\mathrm{r}}_{i} =a^{-{1/2} } \boldsymbol{\mathrm{v}}_{i} ,         
\end{equation} 
where $\boldsymbol{\mathrm{r}}_{i} =a\boldsymbol{\mathrm{x}}_{i} $ is the physical coordinate of particle \textit{i}. 

In this section, the original equation of motion (Eq. \eqref{ZEqnNum157264}) for a comoving system in expanding background is equivalently transformed to Eq. \eqref{ZEqnNum730753} for a transformed system with a constant damping in static background. While two systems are essentially equivalent, analytical solutions can be more accessible in the transformed system for the sake of convenience.

\section{Analytical solutions for TBCM in expanding background}
\label{sec:3}
The two-body gravitational collapse is a fundamental and elementary process. Halos are often created by two-body collapse of two smaller halos with comparable or very different masses (for example, halos merging with a single merger). By this mean, halos pass their mass to larger and larger halos such that two-body gravitational collapse is an elementary step for inverse mass cascade \citep{Xu:2021-Inverse-mass-cascade-mass-function}. Therefore, it should be very instructive to solve a simple two-body collapse model (TBCM) in expanding background.
 
\subsection{Analytical formulation of TBCM model}
\label{sec:3.1}
Solutions are well-known for two-body problem in a static background without damping. Here we focus on the two-body collapse in expanding background. Again, the two-body interaction is assumed to be a general power-law with an exponent \textit{n}. We first analytically solve the TBCM model in transformed system (static background with a constant damping) for convenience. Results can be readily transformed back to the original comoving system. 

\begin{figure}
\includegraphics*[width=\columnwidth]{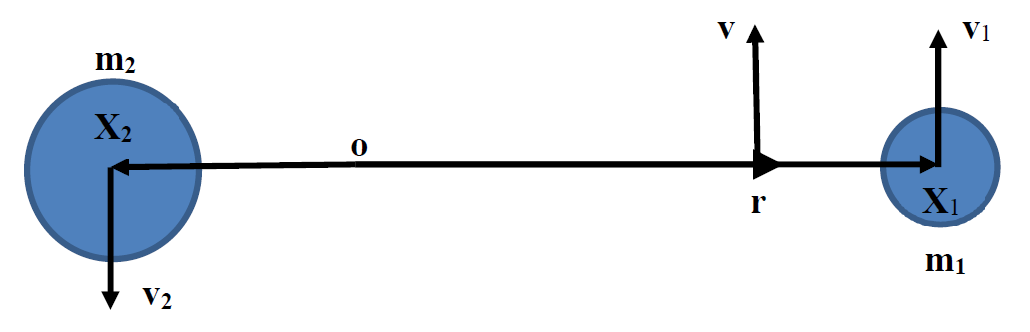}
\caption{Schematic plot of a two-body gravitational collapse in expanding background. The two-body system consists of two masses $m_{1} $ and $m_{2} $ with a separation of 2\textit{r}, where \textbf{\textit{r}} is the displacement vector. Here $\boldsymbol{\mathrm{v}}_{1} $, $\boldsymbol{\mathrm{v}}_{2} $ and $\boldsymbol{\mathrm{v}}$ are the velocities of two masses and the displacement vector, respectively.}
\label{fig:1}
\end{figure}

As shown in Fig. \ref{fig:1}, the two-body system of two masses $m_{1} $ and $m_{2} $ with a separation of 2\textit{r} in expanding background can be equivalently written as (in transformed system from Eq. \eqref{ZEqnNum730753}),
\begin{equation} 
\label{ZEqnNum517916} 
\ddot{\boldsymbol{\mathrm{x}}}_{1} +\frac{H_{0} }{2} \dot{\boldsymbol{\mathrm{x}}}_{1} =\frac{nG_{n} m_{2} }{\left(2r\right)^{1-n} } \cdot \frac{\boldsymbol{\mathrm{r}}}{\left|\boldsymbol{\mathrm{r}}\right|} ,         
\end{equation} 
\begin{equation} 
\label{ZEqnNum191563} 
\ddot{\boldsymbol{\mathrm{x}}}_{2} +\frac{H_{0} }{2} \dot{\boldsymbol{\mathrm{x}}}_{2} =-\frac{nG_{n} m_{1} }{\left(2r\right)^{1-n} } \cdot \frac{\boldsymbol{\mathrm{r}}}{\left|\boldsymbol{\mathrm{r}}\right|} ,         
\end{equation} 
where $\boldsymbol{\mathrm{x}}_{1} $ and $\boldsymbol{\mathrm{x}}_{2} $ are position vectors of two masses and $\boldsymbol{\mathrm{v}}_{j} =\dot{\boldsymbol{\mathrm{x}}}_{j} $ (\textit{j}=1, 2) is the velocity in transformed system with time derivative with respect to \textit{s}.  The displacement vector is defined as $\boldsymbol{\mathrm{r}}={\left(\boldsymbol{\mathrm{x}}_{1} -\boldsymbol{\mathrm{x}}_{2} \right)/2} $ and \textit{r} is the magnitude of vector \textbf{r}. The equation of motion for the center of mass can be obtained by multiplying Eqs. \eqref{ZEqnNum517916} and \eqref{ZEqnNum191563} with $m_{1} $ and $m_{2} $, respectively, and adding them together,
\begin{equation} 
\label{ZEqnNum736305} 
\ddot{\boldsymbol{\mathrm{R}}}+\frac{H_{0} }{2} \dot{\boldsymbol{\mathrm{R}}}=0,          
\end{equation} 
where $\boldsymbol{\mathrm{R}}={\left(m_{1} \boldsymbol{\mathrm{x}}_{1} +m_{2} \boldsymbol{\mathrm{x}}_{2} \right)/\left(m_{1} +m_{2} \right)} $ is the position vector of the center of mass. Similarly, the equation for displacement vector $\boldsymbol{\mathrm{r}}$ can be obtained by subtracting Eq. \eqref{ZEqnNum191563} from Eq. \eqref{ZEqnNum517916},
\begin{equation} 
\label{ZEqnNum365738} 
\begin{split}
&\ddot{\boldsymbol{\mathrm{r}}}+\frac{H_{0} }{2} \dot{\boldsymbol{\mathrm{r}}}=\frac{nG_{n} \left(m_{1} +m_{2} \right)}{2\left(2r\right)^{1-n} } \cdot \frac{\boldsymbol{\mathrm{r}}}{\left|\boldsymbol{\mathrm{r}}\right|},\\
&\textrm{Position vectors can be expressed in terms of \textbf{\textit{r}} and \textbf{\textit{R}} as} \\
&\boldsymbol{\mathrm{x}}_{1} =\boldsymbol{\mathrm{R}}+{2m_{2} \boldsymbol{\mathrm{r}}/\left(m_{1} +m_{2} \right)} =\boldsymbol{\mathrm{R}}+\mu \boldsymbol{\mathrm{r}},\\
&\boldsymbol{\mathrm{x}}_{2} ={\boldsymbol{\mathrm{R}}-2m_{1} \boldsymbol{\mathrm{r}}/\left(m_{1} +m_{2} \right)} =\boldsymbol{\mathrm{R}}-\left(2-\mu \right)\boldsymbol{\mathrm{r}}, 
\end{split}
\end{equation} 
\noindent where $\mu ={2m_{2} /\left(m_{1} +m_{2} \right)}$ is a dimensionless constant. 

We assume a fixed center of mass at the origin \textit{o} (see Fig. \ref{fig:1}) such that $\boldsymbol{\mathrm{R}}=0$ and Eq. \eqref{ZEqnNum736305} is trivial by properly choosing the initial positions and velocities of two masses. The dynamics of the original problem is now reduced to the motion of a point mass subject to gravity and a constant damping ${H_{0} /2} $ (Eq. \eqref{ZEqnNum365738}). This equation exactly mimics a one degree-of-freedom harmonic oscillator, i.e.
\begin{equation}
\label{eq:12}
\ddot{\boldsymbol{\mathrm{r}}}+(c/m) \dot{\boldsymbol{\mathrm{r}}}=-(k/m)  \boldsymbol{\mathrm{r}},
\end{equation}
where $c$ is damping and $k$ is a spring constant. Just like the fundamental role of harmonic oscillator (Eq. \eqref{eq:12}) in dynamics, we will demonstrate the similar role of two-body collapse model plays in self-gravitating collisionless dark matter flow.

Since two-body motion is planar, let's try a general solution for the displacement vector \textbf{\textit{r}} in the \textit{x}-\textit{y} plane, where the Cartesian components of displacement vector \textbf{\textit{r}} and its velocity \textbf{\textit{v}} read
\begin{equation}
x=r\left(s\right)\cos \left(\omega \left(s\right)s\right) \quad \textrm{and} \quad y=r\left(s\right)\sin \left(\omega \left(s\right)s\right),     
\label{ZEqnNum186239}
\end{equation}
\vspace*{-20pt}
\begin{equation} 
\label{ZEqnNum988334} 
\begin{split}
&v_{x} =\dot{x}=\dot{r}\cos \left(\omega s\right)-r\sin \left(\omega s\right)\left(\omega +s\dot{\omega }\right) \\
&\textrm{and}\\ 
&v_{y} =\dot{y}=\dot{r}\sin \left(\omega s\right)+r\cos \left(\omega s\right)\left(\omega +s\dot{\omega }\right). 
\end{split}
\end{equation} 
Both radius $r\left(s\right)$ and frequency term $\omega \left(s\right)$ are functions of time \textit{s}. 

From Eq. \eqref{ZEqnNum365738}, the position and velocity of two masses can be related to that of the displacement vector \textbf{\textit{r}}as
\begin{equation}
\boldsymbol{\mathrm{x}}_{1} =\mu \boldsymbol{\mathrm{r}} \quad \textrm{and} \quad \boldsymbol{\mathrm{x}}_{2} =-\left(2-\mu \right)\boldsymbol{\mathrm{r}}, 
\label{ZEqnNum581329}
\end{equation}
\vspace*{-20pt}
\begin{equation}
\boldsymbol{\mathrm{v}}_{1} =\mu \dot{\boldsymbol{\mathrm{r}}} \quad \textrm{and} \quad \boldsymbol{\mathrm{v}}_{2} =-\left(2-\mu \right)\dot{\boldsymbol{\mathrm{r}}}.      
\label{eq:16}
\end{equation}
\noindent For example, the position ($\boldsymbol{\mathrm{x}}_{1} $) and velocity vectors ($\boldsymbol{\mathrm{v}}_{1} $) of mass $m_{1} $ in \textit{x}-\textit{y} plane can be found as,
\begin{equation}
x_{1} =\mu r\left(s\right)\cos \left(\omega s\right) \quad \textrm{and} \quad y_{1} =\mu r\left(s\right)\sin \left(\omega s\right)      
\label{ZEqnNum435078}
\end{equation}
\vspace*{-20pt}
\begin{equation}
\label{ZEqnNum542855} 
\begin{split}
&v_{x1} =\mu v_{x} =\mu \dot{r}\cos \left(\omega s\right)-\mu r\sin \left(\omega s\right)\left(\omega +s\dot{\omega }\right)\\ 
&\textrm{and}\\ 
&v_{y1} =\mu v_{y} =\mu \dot{r}\sin \left(\omega s\right)+\mu r\cos \left(\omega s\right)\left(\omega +s\dot{\omega }\right). \end{split}
\end{equation} 

The initial positions of two masses $m_{1} $ and $m_{2} $ are set as 
\begin{equation} 
\label{ZEqnNum373431} 
r_{1} \left(s=0\right)=\left|\boldsymbol{\mathrm{x}}_{i1} \right|=\mu r_{i} ,  r_{2} \left(s=0\right)=\left|\boldsymbol{\mathrm{x}}_{i2} \right|=\left(2-\mu \right)r_{i} ,      
\end{equation} 
where $r_{i} =r\left(s=0\right)$ is the magnitude of the displacement vector \textbf{\textit{r}},\textbf{ }$\boldsymbol{\mathrm{x}}_{i1} $ and $\boldsymbol{\mathrm{x}}_{i2} $ are the initial position vectors of two masses. The initial velocities of masses $m_{1} $ and $m_{2} $ can be set as, 
\begin{equation}
v_{x1} \left(s=0\right)=0 \quad \textrm{and} \quad v_{y1} \left(s=0\right)=v_{i1} =\mu v_{i},       
\label{ZEqnNum860465}
\end{equation}
\vspace*{-20pt}
\begin{equation}
v_{x2} \left(s=0\right)=0 \quad \textrm{and} \quad v_{y2} \left(s=0\right)=v_{i2} =-\left(2-\mu \right)v_{i},     
\label{ZEqnNum194669}
\end{equation}

\noindent where $v_{i} $ is the initial velocity of the displacement vector \textbf{\textit{r}} in \textit{y} direction. These initial conditions satisfy a zero linear momentum with $m_{1} v_{i1} +m_{2} v_{i2} =0$. Obviously, $\boldsymbol{\mathrm{R}}=0$ is a trivial solution for Eq. \eqref{ZEqnNum736305} with these initial conditions (Eqs. \eqref{ZEqnNum373431}, \eqref{ZEqnNum860465}, \eqref{ZEqnNum194669}). 

A special case is that the initial speed $v_{i}$ of vector \textbf{\textit{r}} satisfies
\begin{equation} 
\label{ZEqnNum443254} 
v_{i} =v_{ri} =\sqrt{\frac{-nG_{n} r_{i} }{\left(2r_{i} \right)^{1-n} } \frac{m_{1} +m_{2} }{2} } ,          
\end{equation} 
where the corresponding speeds of two masses $v_{i1}$ and $v_{i2}$ are
\begin{equation}
\frac{v_{i1}^{2} }{\left|\boldsymbol{\mathrm{x}}_{i1} \right|} =\frac{-nG_{n} m_{2} }{\left(2r_{i} \right)^{1-n} } \quad \textrm{and}  \quad \frac{v_{i2}^{2} }{\left|\boldsymbol{\mathrm{x}}_{i2} \right|} =\frac{-nG_{n} m_{1} }{\left(2r_{i} \right)^{1-n} }.     
\label{ZEqnNum318624}
\end{equation}

\noindent Here $v_{ri} $ is the circling velocity of the displacement vector \textbf{\textit{r}} if there is no damping. For this special case, the two-body system is stable with both masses circling around the center of mass if the background is static ($H_{0} =0$). More specifically, the initial system is in a virial equilibrium ($2KE-nPE=0$) with
\begin{equation} 
\label{ZEqnNum577609} 
v_{i}^{2} =v_{ri}^{2} =\alpha _{s} \frac{G_{n} \left(m_{1} +m_{2} \right)}{r_{i}^{-n} } =\alpha _{s} \frac{G_{n} M}{r_{i}^{-n} } ,        
\end{equation} 
where constant $\alpha _{s} ={-n/2^{2-n} } $ and $M=m_{1} +m_{2} $ is the total mass of the system. 

Without loss of generality, we will try to solve the two-body collapse problem with an arbitrary initial velocity $v_{i} $ for displacement vector \textbf{\textit{r}}. Substituting the assumed solution (Eq. \eqref{ZEqnNum186239}) into Eq. \eqref{ZEqnNum365738} gives rise to two coupled equations for two unknown functions: radius $r\left(s\right)$ and frequency $\omega \left(s\right)$,
\begin{equation} 
\label{ZEqnNum804764} 
\ddot{r}+\underbrace{\frac{H_{0} }{2} }_{1}\dot{r}-\underbrace{\frac{nG_{n} \left(m_{1} +m_{2} \right)}{2\left(2r\right)^{1-n} } }_{2}=\underbrace{r\left(\omega +s\dot{\omega }\right)^{2} }_{3}=r\left(\frac{\partial \left(\omega s\right)}{\partial s} \right)^{2} ,     
\end{equation} 
\begin{equation} 
\label{ZEqnNum294824} 
\frac{\dot{r}}{r} =-\frac{1}{2} \left[\frac{\partial \ln \left(\omega +s\dot{\omega }\right)}{\partial s} +\frac{H_{0} }{2} \right].        
\end{equation} 
with initial conditions,
\begin{equation}
\left. r\right|_{s=0} =r_{i} \quad \textrm{and} \quad \left. \left(\frac{\partial r}{\partial s} \right)\right|_{s=0} =0.  \label{ZEqnNum305720}
\end{equation}

Three forces contribute to the equation of motion for $r\left(s\right)$ in Eq. \eqref{ZEqnNum804764}, i.e. the damping force (term 1), the gravitational force (term 2), and the frequency force (term 3). Term 3 (frequency force) origins from the angular momentum as we will show in Eq. \eqref{ZEqnNum332742}. The competition among three forces dominates the evolution of $r\left(s\right)$. Let's now introduce a frequency function as
\begin{equation} 
\label{ZEqnNum553834} 
F\left(s\right)=\left(\omega +s\dot{\omega }\right)^{{-1/2} } =\left(\frac{\partial \left(\omega s\right)}{\partial s} \right)^{{-1/2} } .         
\end{equation} 
The radius $r\left(s\right)$ can be obtained by solving Eq. \eqref{ZEqnNum294824},
\begin{equation} 
\label{ZEqnNum197153} 
r\left(s\right)=\left(r_{i} v_{i} \right)^{{1/2} } F\left(s\right)\exp \left(-\frac{1}{4} H_{0} s\right).        
\end{equation} 

A single equation for radius $r\left(s\right)$ can be easily obtained by substitution of Eq. \eqref{ZEqnNum197153} for frequency function $F\left(s\right)$ into Eq. \eqref{ZEqnNum804764},
\begin{equation} 
\label{ZEqnNum332742} 
\ddot{r}+\frac{H_{0} }{2} \dot{r}-\frac{nG_{n} \left(m_{1} +m_{2} \right)}{2\left(2r\right)^{1-n} } =\frac{\left(r_{i} v_{i} \right)^{2} }{r^{3} } \exp \left(-H_{0} s\right).      
\end{equation} 
The frequency force (term 3 in Eq. \eqref{ZEqnNum804764}) is now related to the initial angular momentum ($r_{i} v_{i} $ on the right hand side (RHS)) and is exponentially decaying with time \textit{s}. Complete solution of Eq. \eqref{ZEqnNum332742} depends on five parameters, i.e. the exponent $n$, damping $H_{0} $, initial conditions $r_{i} $ and $v_{i} $, and system mass $M=m_{1} +m_{2} $. This equation mimics the spherical collapse model (SCM) but with a non-zero angular momentum on RHS. Comparison is discussed in Section \ref{sec:4.3}. 

However, Eq. \eqref{ZEqnNum332742} is complex to solve analytically. Here we take a different route by directly solving the frequency function $F\left(s\right)$ (instead of $r\left(s\right)$ in Eq. \eqref{ZEqnNum332742}) , where five parameters can be grouped and significantly reduced to exponent \textit{n} and two dimensionless numbers (Eqs. \eqref{ZEqnNum129548} and \eqref{ZEqnNum629408}). Equations \eqref{ZEqnNum804764} and \eqref{ZEqnNum294824} are first combined and rewritten in terms of the frequency function $F\left(s\right)$, 
\begin{equation} 
\label{ZEqnNum928746} 
\frac{\ddot{r}}{r} +\frac{H_{0} }{2} \frac{\dot{r}}{r} +\frac{v_{ri}^{2} }{r^{2} } \left(\frac{r}{r_{i} } \right)^{n} =F^{-4} \left(s\right),        
\end{equation} 
\begin{equation} 
\label{ZEqnNum165887} 
\frac{\dot{r}}{r} =\frac{1}{F\left(s\right)} \frac{\partial F}{\partial s} -\frac{H_{0} }{4} .          
\end{equation} 
With the identity 
\begin{equation} 
\label{eq:33} 
\frac{\ddot{r}}{r} =\frac{\partial \left({\dot{r}/r} \right)}{\partial s} +\left(\frac{\dot{r}}{r} \right)^{2} ,           
\end{equation} 
substitution of Eq. \eqref{ZEqnNum165887} into Eq. \eqref{ZEqnNum928746} leads to a single equation for frequency function $F\left(s\right)$ (no first order derivative involved):
\begin{equation} 
\label{ZEqnNum502680} 
\frac{\partial ^{2} F}{\partial s^{2} } =\underbrace{\frac{H_{0}^{2} }{16} F\left(s\right)}_{1}-\underbrace{\gamma _{s} \left(\frac{v_{i} }{r_{i} } \right)^{1+{n/2} } F^{n-1} \left(s\right)\exp \left(-\frac{n-2}{4} H_{0} s\right)}_{2}+\underbrace{F^{-3} \left(s\right)}_{3} 
\end{equation} 
with initial conditions:
\begin{equation}
F\left(s=0\right)=\left(\frac{r_{i} }{v_{i} } \right)^{{1/2} } \quad \textrm{and} \quad \left. \frac{\partial F}{\partial s} \right|_{s=0} =\frac{H_{0} }{4} \left(\frac{r_{i} }{v_{i} } \right)^{{1/2} },  
\label{ZEqnNum124395}
\end{equation}

\noindent from (Eq. \eqref{ZEqnNum305720}), where $\gamma _{s} =\left({v_{ri} /v_{i} } \right)^{2} $ is a dimensionless number indicating how far the initial system is away from virial equilibrium (the special case in (Eq. \eqref{ZEqnNum443254}). $\gamma _{s} =1$ corresponds to the special case with initial system in virial equilibrium. 

With function $F\left(s\right)$ fully determined by the Eq. \eqref{ZEqnNum502680} and initial condition in \eqref{ZEqnNum124395}, the radius $r\left(s\right)$ and frequency $\omega \left(s\right)$ can be solved subsequently using Eqs. \eqref{ZEqnNum197153} and \eqref{ZEqnNum553834}. Similarly, three terms (1, 2 and 3) on the RHS of Eq. \eqref{ZEqnNum502680}, i.e. the damping force, the gravitational force, and the frequency force (from angular momentum), contribute to the evolution of $F\left(s\right)$. 

\subsection{Numerical solutions and three distinct regimes for TBCM}
\label{sec:3.2}
\begin{figure}
\includegraphics*[width=\columnwidth]{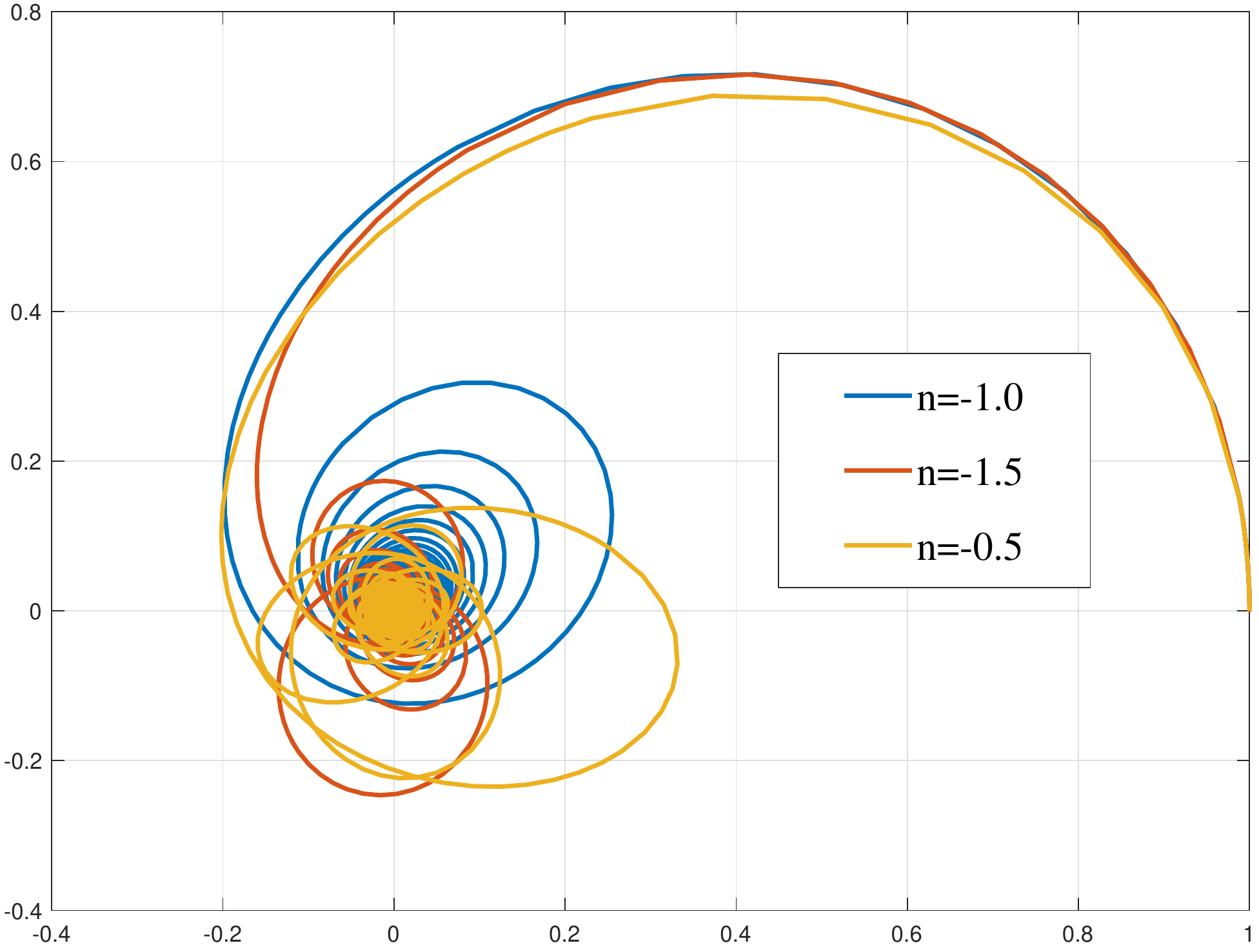}
\caption{The trajectory of the displacement vector \textbf{\textit{r}} in \textit{x}-\textit{y} plane for three different \textit{n} with $H_{0} =0.4$, $G_{n} M=1$, $r_{i} =1$ and $\gamma _{s} =1$ (or $v_{i} =v_{ri} $), i.e. the initial system is in virial equilibrium. The trajectory becomes very complex for systems with different potential exponents \textit{n} and a nonzero damping ($H_{0} \ne 0$ stands for expanding background).}
\label{fig:2}
\end{figure}

Exact solution of highly nonlinear Eq. \eqref{ZEqnNum502680} is still not available in a closed form. However, numerical solutions can be easily obtained. Figure \ref{fig:2} shows typical trajectories of displacement vector \textbf{\textit{r}} in \textit{x}-\textit{y} plane for three different \textit{n }= -0.5, -1.0, and -1.5. The trajectories are for the gravitational collapse of two masses in a transformed system. Initial systems are in virial equilibrium, where displacement vector \textbf{\textit{r}} simply circles around the origin if $H_{0} =0$. The trajectory becomes very complex for systems with different potential exponent \textit{n} and a nonzero damping ($H_{0} \ne 0$ for expanding background). 

\begin{figure}
\includegraphics*[width=\columnwidth]{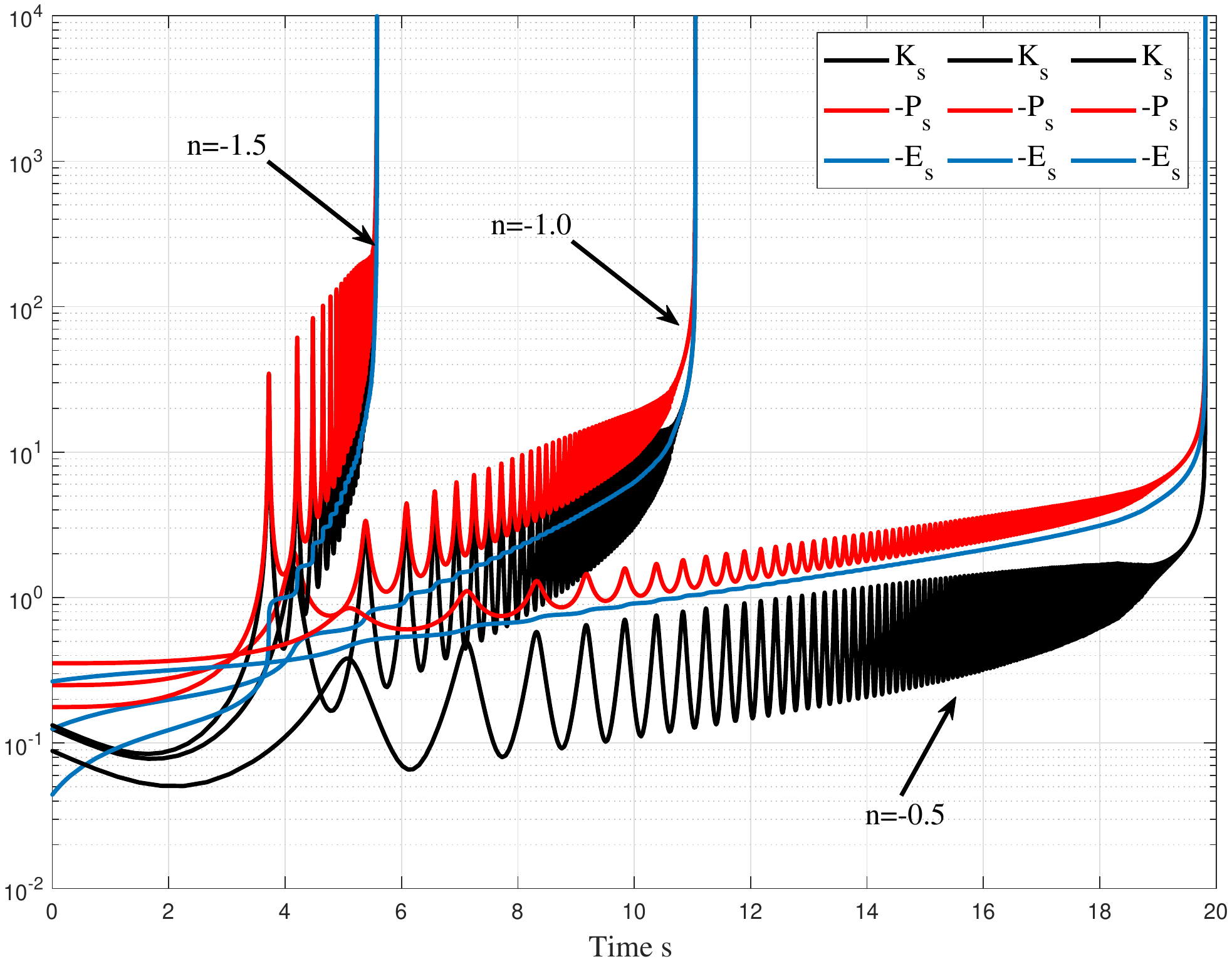}
\caption{The temporal evolution of energy for a two-body gravitational collapse for three different exponents \textit{n }with $H_{0} =0.4$,$G_{n} M=1$, $r_{i} =1$ and $\gamma _{s} =1$ (or $v_{i} =v_{ri} $). Both kinetic and potential energy ($K_{s} $ and $P_{s} $) vibrate around their mean values before the final collapse. The oscillation cancels out for the total energy $E_{s} =K_{s} +P_{s} $. A smaller exponent \textit{n} tends to have a longer time span of oscillation and smaller amplitude of oscillation.}
\label{fig:3}
\end{figure}

Figure \ref{fig:3} plots the time evolution of specific kinetic, potential, and total energy for the same three cases in Fig. \ref{fig:2}. Both kinetic and potential energies of two-body system ($K_{s} $ and $P_{s} $) vibrate around their mean values before the final collapse. The oscillation cancels out for total energy $E_{s} =K_{s} +P_{s} $, which is relatively smooth. A smaller exponent \textit{n} tends to have a longer time span of oscillation and smaller oscillation amplitude. More detailed discussion of energy evolution and their solutions is presented in Section \ref{sec:3.6}. 

\begin{figure}
\begin{tabular}{p{1.6in}p{1.6in}} 
\includegraphics*[width=1.7in, height=1.3in]{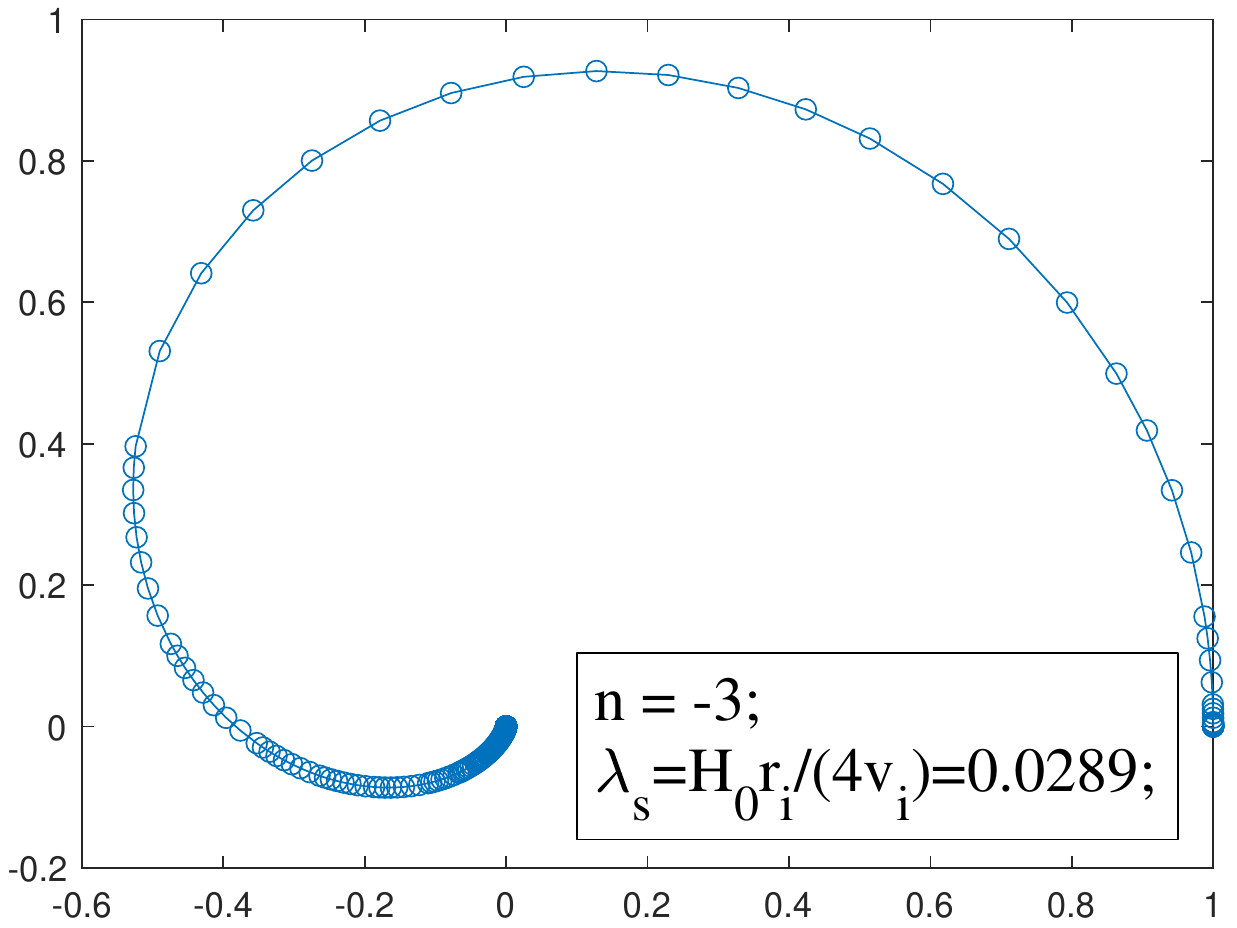} & \includegraphics*[width=1.7in, height=1.3in]{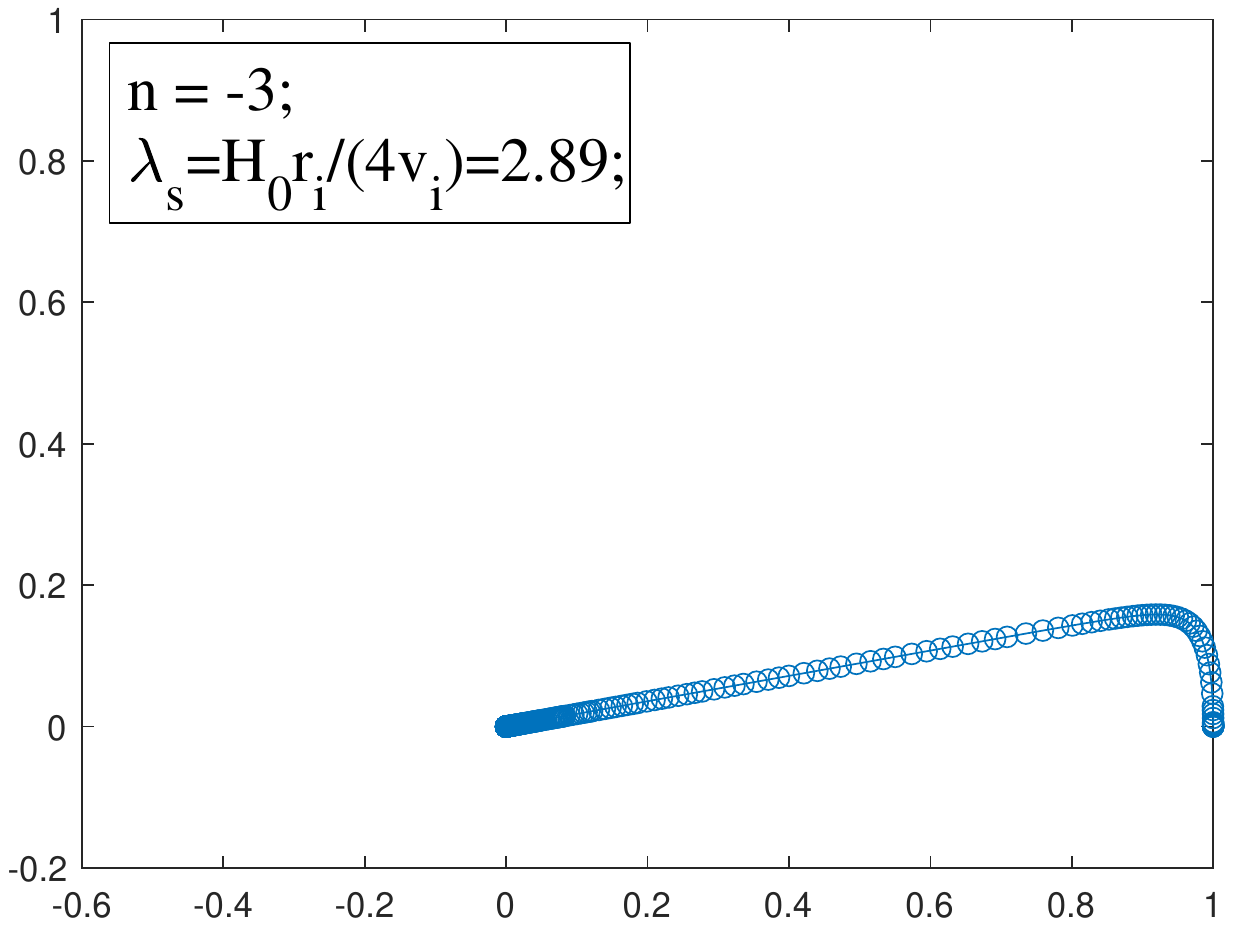} \\  
\includegraphics*[width=1.7in, height=1.3in]{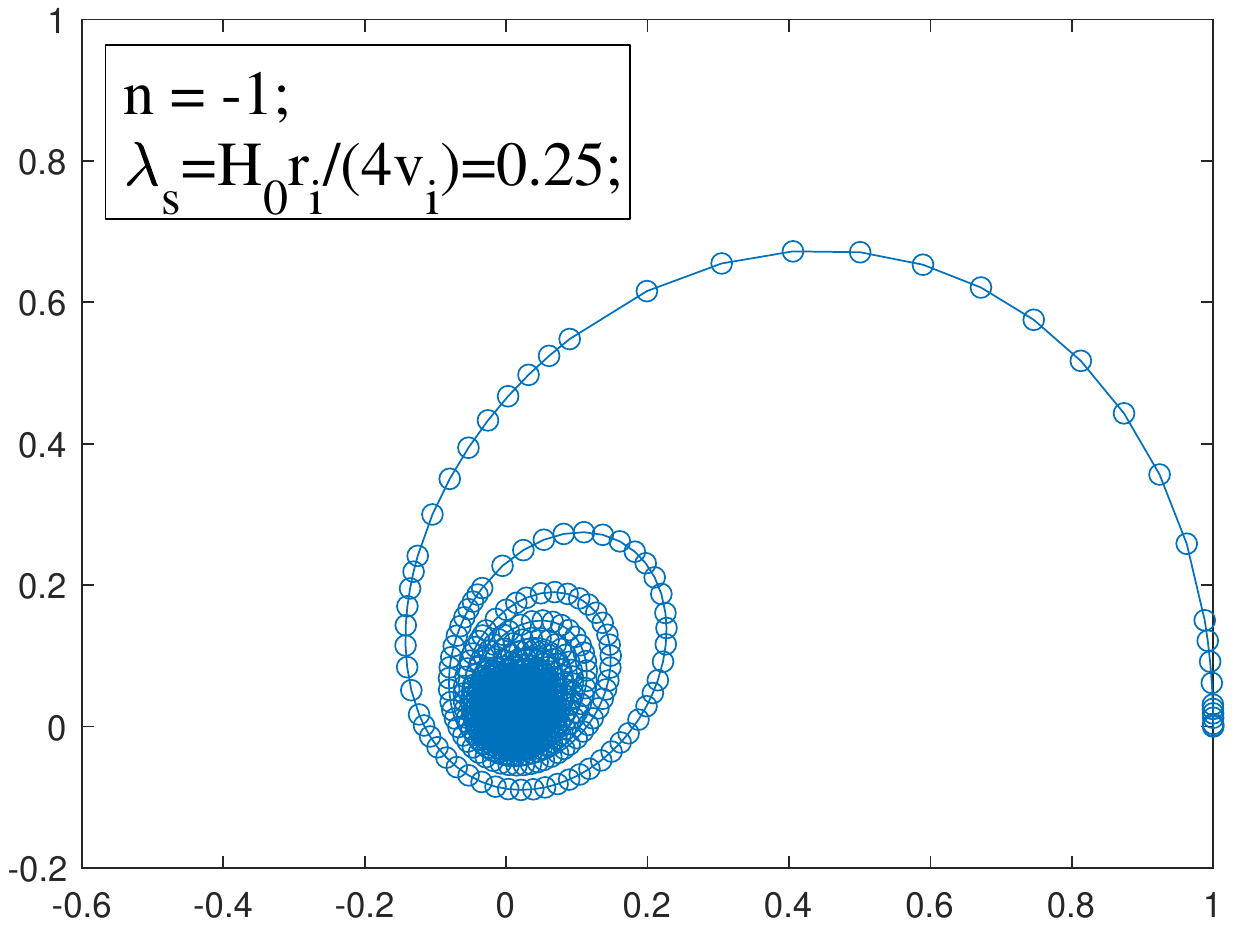} & \includegraphics*[width=1.7in, height=1.3in]{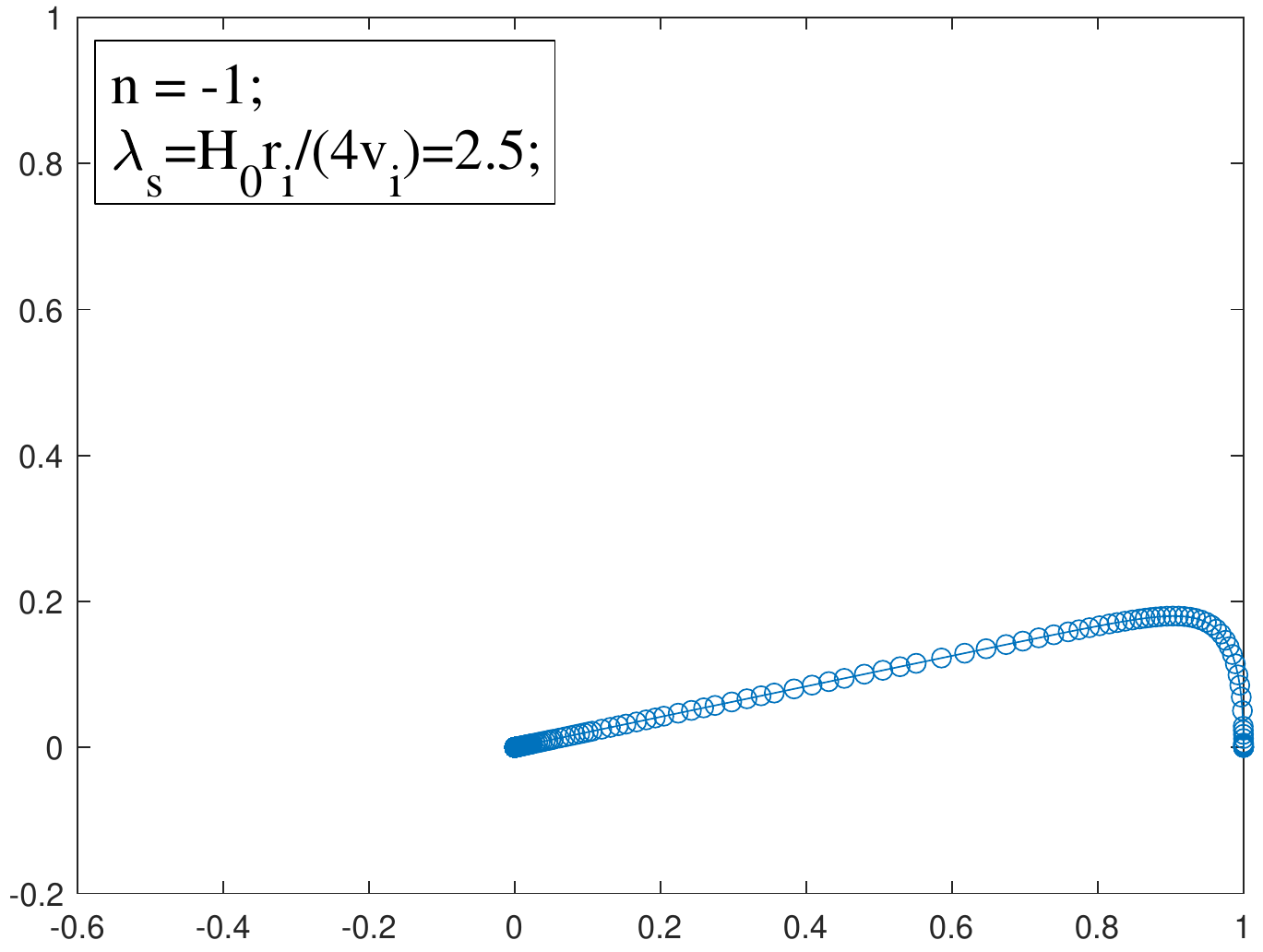} \\  
\end{tabular}
\caption{Four typical trajectories for different combinations of parameter $\lambda _{s} $ and potential exponent $n$ with $\gamma _{s} =1$. The oscillatory motion only exists for small $\lambda _{s} $ (weak damping) and$-2<n<0$. The critical value of $\lambda _{s} $ for an oscillatory motion will be identified.}
\label{fig:4}
\end{figure}

Figure \ref{fig:4} presents typical trajectories for four different scenarios, depending on a dimensionless number ${\lambda _{s} =H_{0} r_{i} /\left(4v_{i} \right)} $ and the exponent \textit{n}. All scenarios have $\gamma _{s} =1$ or $v_{i} =v_{ri} $, i.e. the special case considered in Eq. \eqref{ZEqnNum443254}. The periodic motion only exists for small $\lambda _{s} $ and$-2<n<0$. The dimensionless number $\lambda _{s} $ quantifies the competition between gravity and damping, while the ratio $\gamma _{s} =\left({v_{ri} /v_{i} } \right)^{2} $ quantifies the competition between gravity and angular momentum. There exists a critical value of $\lambda _{s} $ for the existence of periodic motion that we will identify later. This exactly mimics the critical damping $c_s=2\sqrt{km}$ for harmonic oscillator in Eq. \eqref{eq:12}, above which damping is dominant to eliminate the periodic motion (overdamped system). 

\begin{figure}
\includegraphics*[width=\columnwidth]{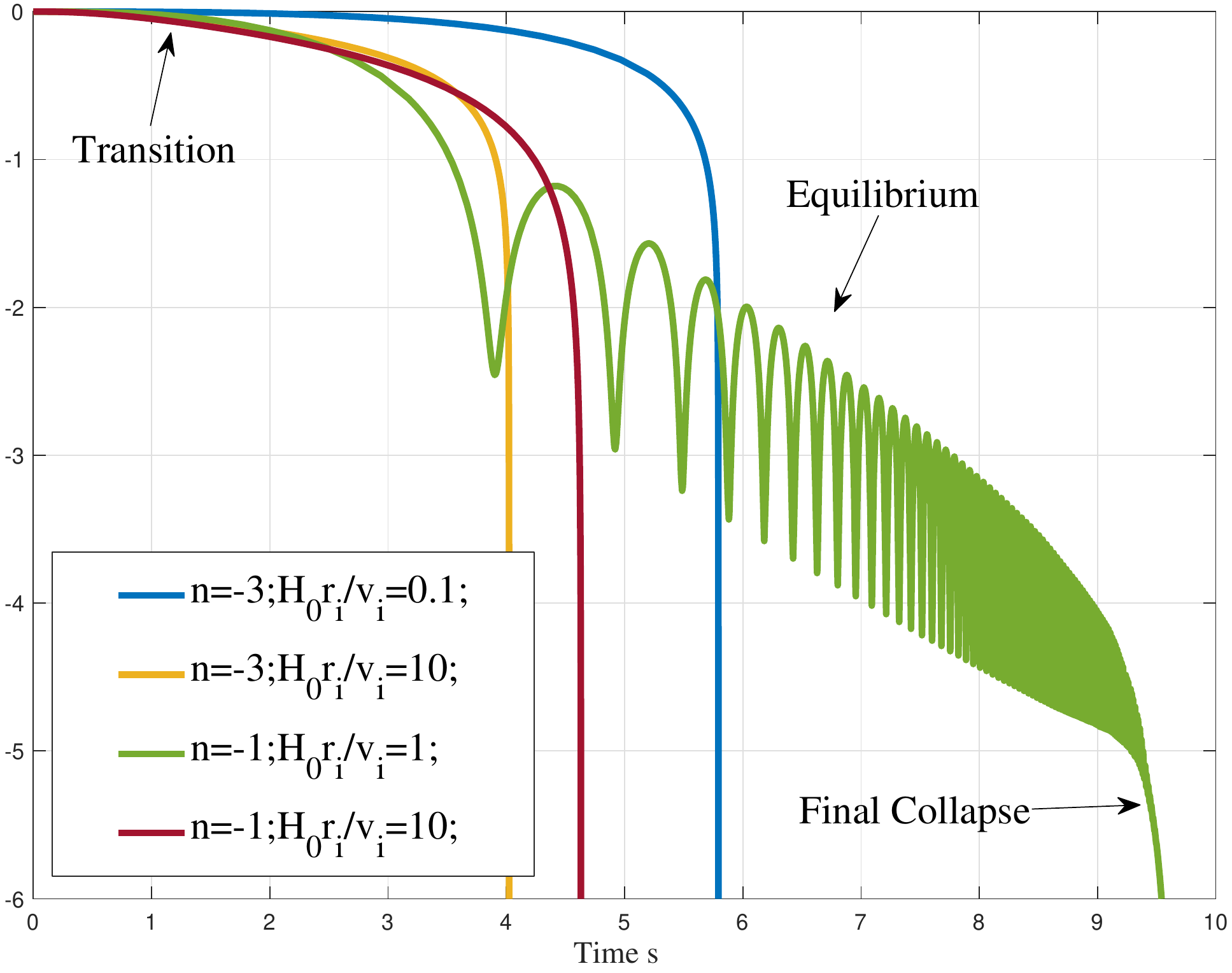}
\caption{The temporal evolution of radius function $r\left(s\right)$ with time \textit{s} for four cases presented in Fig. \ref{fig:4}. For an equilibrium collapse with oscillating motion (under damped), three distinct ranges can be identified, an initial transition range dominated by the damping force, an equilibrium range dominated by the competition between the gravitational and the frequency forces, and a final collapse. The equilibrium collase only exists for weak damping with a small $\lambda _{s} $, $-2<n<0$, and $\gamma _{s} \approx 1$.}
\label{fig:5}
\end{figure}

Figure \ref{fig:5} shows the time variation of the radius function $r\left(s\right)$ of displacement vector for the same four scenarios in Fig. \ref{fig:4}. Three distinct regimes can be identified for an equilibrium collapse (green line), i.e. an initial transitional range dominated by the damping force, an equilibrium range dominated by the competition between the gravitational and the frequency forces, and a final collapse. System spends most time in the equilibrium range if an oscillating motion exists, which corresponds to the statistically steady state in SG-CFD.  

Term 1 (damping) on the RHS of Eq. \eqref{ZEqnNum502680} can be dominant over the other two terms initially. The solution of $F\left(s\right)$ for the transition range can be found as,
\begin{equation}
F\left(s\right)=\left(\frac{r_{i} }{v_{i} } \right)^{{1/2} } \exp \left(\frac{H_{0} s}{4} \right) \quad \textrm{and} \quad r\left(s\right)=r_{i}.
\label{ZEqnNum288261}
\end{equation}

\noindent Since term 1 (damping) is dominant at the transition range, we have (from Eq. \eqref{ZEqnNum502680}),
\begin{equation} 
\label{ZEqnNum308379} 
\frac{H_{0}^{2} }{16} F\left(s_{t} \right)=\left|\gamma _{s} \left(\frac{v_{i} }{r_{i} } \right)^{1+{n/2} } F^{n-1} \left(s_{t} \right)\exp \left(-\frac{n-2}{4} H_{0} s_{t} \right)-F^{-3} \left(s_{t} \right)\right| 
\end{equation} 
to define a transition time $s_{t} $. After substitution of Eq. \eqref{ZEqnNum288261} into Eq. \eqref{ZEqnNum308379}, we have
\begin{equation} 
\label{eq:38} 
\lambda _{s}^{2} =\left|\gamma _{s} -\exp \left(-H_{0} s_{t} \right)\right|,          
\end{equation} 
where the dimensionless number $\lambda _{s} $ is defined as $\lambda _{s} ={H_{0} r_{i} /\left(4v_{i} \right)} $ and the transition time $s_{t} $ is dependent on $\lambda _{s} $ and $\gamma _{s} $. Damping force is dominant for $t<s_{t} $.

For equilibrium range, term 2 (gravitational force) approximately balances the term 3 (the frequency force) which leads to a mean frequency function $F_{m} \left(s\right)$ from Eq. \eqref{ZEqnNum502680},
\begin{equation} 
\label{ZEqnNum527908} 
F_{m} \left(s\right)=\gamma _{s}^{-{1/\left(2+n\right)} } \left(\frac{r_{i} }{v_{i} } \right)^{{1/2} } \exp \left(-\frac{2-n}{2+n} \cdot \frac{H_{0} s}{4} \right).       
\end{equation} 
The actual solution $F\left(s\right)$ vibrates around the mean solution $F_{m} \left(s\right)$. The mean solutions for the radius and frequency can be found using Eqs. \eqref{ZEqnNum197153} and \eqref{ZEqnNum553834}, 
\begin{equation} 
\label{ZEqnNum758036} 
r_{m} \left(s\right)=\gamma _{s}^{-{1/\left(2+n\right)} } r_{i} \exp \left(-\frac{H_{0} s}{2+n} \right),        
\end{equation} 
\begin{equation} 
\label{ZEqnNum870001} 
\omega _{m} \left(s\right)=\frac{1}{2\lambda _{s} s} \frac{2+n}{2-n} \gamma _{s}^{{2/\left(2+n\right)} } \exp \left(\frac{2-n}{2+n} \cdot \frac{H_{0} s}{2} \right).       
\end{equation} 
Actual radius and frequency solutions should also vibrate about mean solutions (Fig. \ref{fig:5}). 

\subsection{Free fall collapse and free fall time in expanding background}
\label{sec:3.3}
The free fall time is the characteristic time it takes for two-body to collapse under their own gravity. The TBCM model can be used to estimate the free fall time in expanding background. For small initial velocity with $v_{i} \to 0$ (vanishing angular momentum) or large exponent \textit{n }($v_{ri} \to \infty $ from Eq. \eqref{ZEqnNum443254}), the parameter $\gamma _{s} =\left({v_{ri} /v_{i} } \right)^{2} \to \infty$, i.e. a zero angular momentum. This is the free fall of a test particle from rest at an initial distance of $r_{i} $ with a fixed damping. Term 2 (gravitational force) in Eq. \eqref{ZEqnNum502680} should be dominant for free fall and the solution of $F\left(s\right)$ is approximately a parabolic function without oscillatory motion. From Eq. \eqref{ZEqnNum502680} and initial condition in Eq. \eqref{ZEqnNum124395}, equation for $F\left(s\right)$ reads
\begin{equation} 
\label{ZEqnNum807002} 
\frac{\partial ^{2} F}{\partial s^{2} } \approx -\gamma _{s} \left(\frac{v_{i} }{r_{i} } \right)^{1+{n/2} } F^{n-1} \left(s=0\right)=-\gamma _{s} \left(\frac{v_{i} }{r_{i} } \right)^{{3/2} } .      
\end{equation} 
With initial conditions in Eq. \eqref{ZEqnNum124395}, the solution of Eq. \eqref{ZEqnNum807002} is   
\begin{equation} 
\label{eq:43} 
F\left(s\right)\approx \left(\frac{r_{i} }{v_{i} } \right)^{{1/2} } \left[-\frac{1}{32} \left(\frac{H_{0} }{\lambda _{si} } \right)^{2} s^{2} +\frac{1}{4} H_{0} s+1\right],       
\end{equation} 
where parameter (quantifies the competition of gravity with damping for free fall collapse) 
\begin{equation} 
\label{ZEqnNum252248} 
\lambda _{si} =\frac{H_{0} r_{i} }{4v_{ri} } =\lambda _{s} \gamma _{s}^{-{1/2} } =\frac{\lambda _{s} v_{i} }{v_{ri} } =H_{0} \sqrt{\frac{2^{-n-2} r_{i}^{2-n} }{-nG_{n} \left(m_{1} +m_{2} \right)} } .        
\end{equation} 
The radius function $r(s)$ can be found from Eq. \eqref{ZEqnNum197153},
\begin{equation} 
\label{eq:45} 
r\left(s\right)\approx r_{i} \exp \left(-\frac{1}{4} H_{0} s\right)\left[-\frac{1}{32} \left(\frac{H_{0} }{\lambda _{si} } \right)^{2} s^{2} +\frac{1}{4} H_{0} s+1\right].      
\end{equation} 
The final collapse (free fall) time $s_{c} $ in time scale \textit{s} can be estimated by setting $r\left(s=s_{c} \right)=0$,
\begin{equation} 
\label{ZEqnNum216491} 
s_{c} \approx \frac{4\lambda _{si}^{2} }{H_{0} } \left[1+\sqrt{1+\frac{2}{\lambda _{si}^{2} } } \right].         
\end{equation} 

For small $\lambda _{si}^{} $ ($r_{i} \to 0$, or weak damping $H_{0} \to 0$, or $M\to \infty $) in Eq. \eqref{ZEqnNum252248}, $s_{c} $ is essentially the free fall time in static background without damping (from Eq. \eqref{ZEqnNum216491}),  
\begin{equation} 
\label{ZEqnNum657621} 
s_{c} \approx s_{c1} =4\sqrt{2} \frac{\lambda _{si} }{H_{0} } =\sqrt{\frac{2^{3-n} r_{i}^{2-n} }{-nG_{n} \left(m_{1} +m_{2} \right)} } =\frac{\sqrt{2} }{2\pi } T_{ri}  ,      
\end{equation} 
which is independent of damping $H_{0}$. Here $T_{ri}$ is the orbital period,
\begin{equation} 
\label{eq:48} 
T_{ri} =\frac{2\pi r_{i} }{v_{ri} } =\frac{2\pi \left(2r_{i} \right)^{1-{n/2} } }{\sqrt{-nG_{n} \left(m_{1} +m_{2} \right)} } .         
\end{equation} 
Specifically, for $n=-1$, we have 
\begin{equation} 
\label{eq:49} 
s_{c1} =\frac{4r_{i}^{{3/2} } }{\sqrt{G\left(m_{1} +m_{2} \right)} } =\frac{\sqrt{2} }{2\pi } T_{ri} ,         
\end{equation} 
which well approximates the exact free fall time $s_{ce} $ in static background without damping, where
\begin{equation} 
\label{ZEqnNum306004} 
s_{ce} =\frac{\pi r_{i}^{{3/2} } }{\sqrt{G\left(m_{1} +m_{2} \right)} } =\frac{\sqrt{2} }{8} T_{ri}  
\end{equation} 
is the exact free fall time in static background. Note that $s_{c1} $ is analytically obtained from the approximation Eq. \eqref{ZEqnNum807002} and cannot fully reduce to the exact free fall time $s_{ce}$. 

While for large $\lambda _{si}^{} $ ($r_{i} \to \infty $, or strong damping $H_{0} \to \infty $, or $M\to 0$) in Eq. \eqref{ZEqnNum252248}, the free fall time is proportional to $H_{0}$, 
\begin{equation} 
\label{ZEqnNum992618} 
s_{c} \approx s_{c2} =8\frac{\lambda _{si}^{2} }{H_{0} } =\frac{H_{0} r_{i}^{2} }{2v_{ri}^{2} } =\frac{H_{0} }{8\pi ^{2} } T_{ri}^{2} =\frac{H_{0} 2^{1-n} r_{i}^{2-n} }{-nG_{n} \left(m_{1} +m_{2} \right)} .       
\end{equation} 

\begin{figure}
\includegraphics*[width=\columnwidth]{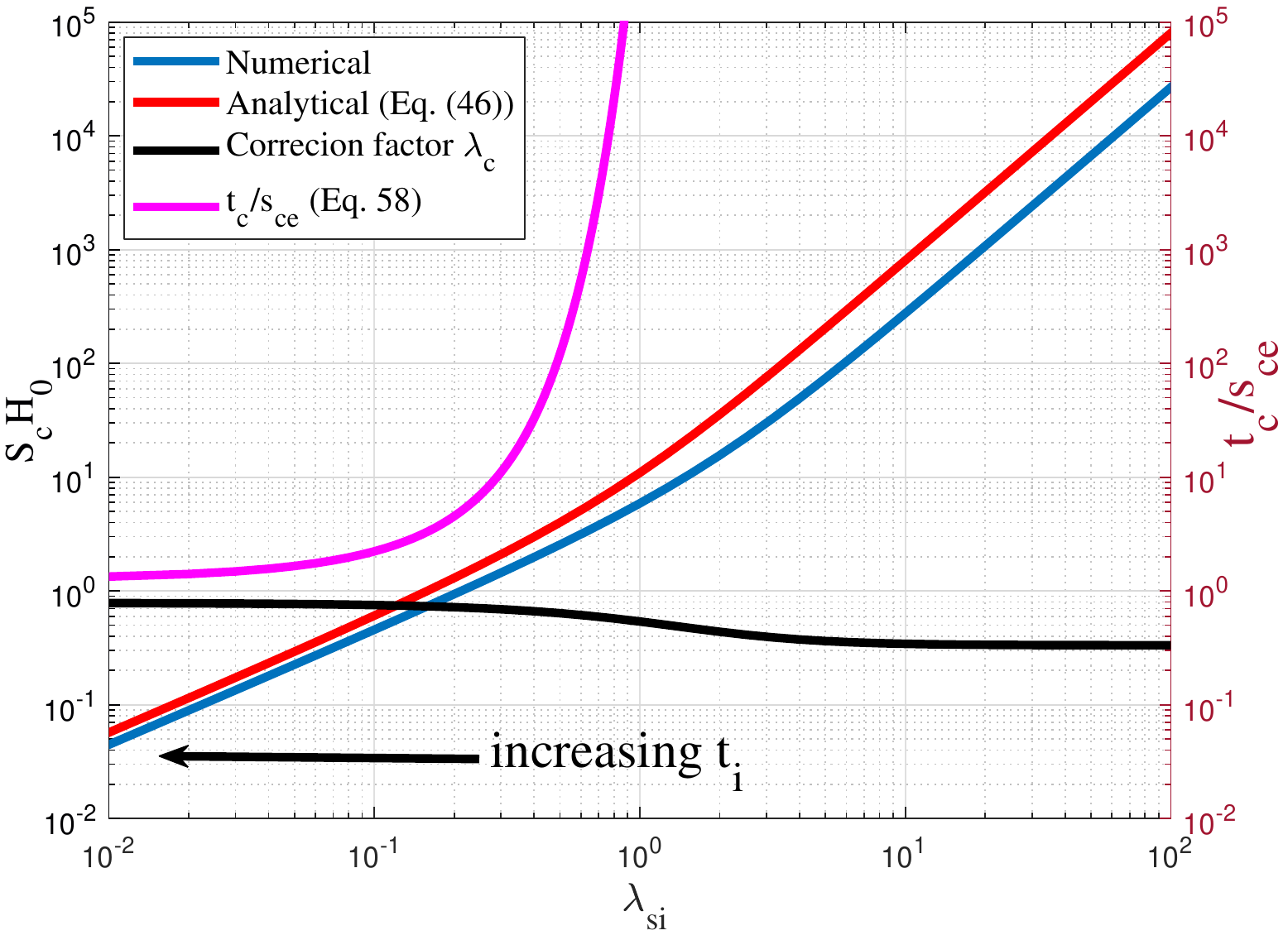}
\caption{The variation of free fall time $s_{c} $ (normalized by $H_{0} $) with dimensionless number $\lambda _{si}^{} $. The comparison between numerical solution by solving the original equation of motion and the analytical approximation is also presented. The ratio between two is plotted as the correction factor $\lambda _{c} $. The free fall time $s_{c} $(in transformed system with static background and fixed damping) is proportional to damping $H_{0} $ for large $\lambda _{si}^{} $. The variation of the physical free fall time $t_{c} $ (normalized by the free fall time $s_{ce} $ that is for static background and no damping) with $\lambda _{si}^{} $ is also plotted on the right axis, which increases with $\lambda _{si}^{} $. The two-body system starting to collapse at an earlier time will have a longer free fall time. Time $t_{c} $ approaches $s_{ce} $ when  $s_{ce} $ is small (small separation or large mass) or when collapse time $t_{i} $ approaches $t_{0} $.}
\label{fig:6}
\end{figure}

The critical value between two regimes can be obtained from Eqs. \eqref{ZEqnNum657621} and \eqref{ZEqnNum992618} with $s_{c1} =s_{c2} $, where $\lambda _{si}^{} ={\sqrt{2} /2} $. The free fall time from Eq. \eqref{ZEqnNum216491} approximates the true free fall time in transformed system. Figure \ref{fig:6} plots the variation of free fall time (normalized by $H_{0} $) with the dimensionless number $\lambda _{si}^{} $. The comparison between the numerical solution by solving the original Equation (Eqs. \eqref{ZEqnNum517916} and \eqref{ZEqnNum191563}) and approximation Eq. \eqref{ZEqnNum216491} justifies a correction factor $\lambda _{c} $, 
\begin{equation}
\begin{split}
&s_{c} =4\lambda _{c} \frac{\lambda _{si}^{2} }{H_{0} } \left[1+\sqrt{1+\frac{2}{\lambda _{si}^{2} } } \right],\\ 
&\textrm{where the correction factor}\\
&\lambda _{c} =\frac{\pi }{4} \quad \textrm{for} \quad \lambda _{si}^{} \to 0 \quad \textrm{and} \quad \lambda _{c} =\frac{1}{3} \quad \textrm{for} \quad \lambda _{si}^{} \to \infty.
\label{ZEqnNum863096}
\end{split}
\end{equation}

Note that $s_{c} $ is the free fall time in transformed system. To transform it back to the original comoving system, the relation between time scales \textit{t} and \textit{s} (${ds/dt} =a^{{-3/2} } $) is 
\begin{equation}
s=t_{0} \ln \left({t/t_{i} } \right) \quad \textrm{and} \quad t_{e} =t_{i} \exp \left({s_{c} /t_{0} } \right),    
\label{ZEqnNum185878}
\end{equation}

\noindent where $t_{i} =a_{i}^{{3/2} } t_{0} $ and $t_{e} $ are the start and end of a two-body free fall in physical time \textit{t}, $a_{i} $ is the scale factor at initial time $t_{i} $. Here $t_{0} $ is the present physical time with $H_{0} t_{0} ={2/3} $. The free fall time ($t_{c} =t_{e} -t_{i} $) for a two-body system to fully collapse in expanding background is,
\begin{equation} 
\label{ZEqnNum497055} 
t_{c} =t_{i} \left(\exp \left(\frac{s_{c} }{t_{0} } \right)-1\right)\approx \frac{t_{i} s_{c} }{t_{0} } .         
\end{equation} 

Let's consider a two-body system with an initial separation of $2r_{yi}^{} $ in physical coordinates, the exact free fall time for such a two-body system in static background without damping should be (same as Eq. \eqref{ZEqnNum306004}, but in a physical coordinate $r_{yi} =a_{i} r_{i} $),
\begin{equation} 
\label{eq:55} 
s_{ce} =\frac{\pi r_{yi}^{{3/2} } }{\sqrt{G\left(m_{1} +m_{2} \right)} } .          
\end{equation} 

For $n=-1$, the dimensionless $\lambda _{si}^{} $ can be rewritten in terms of the ratio ${s_{ce} /t_{i} } $ from Eq. \eqref{ZEqnNum252248}, 
\begin{equation} 
\label{ZEqnNum156295} 
\lambda _{si} =H_{0} \frac{\left({r_{yi} /a_{i} } \right)^{{3/2} } }{\sqrt{2G_{n} \left(m_{1} +m_{2} \right)} } =\frac{\sqrt{2} }{3\pi } \frac{s_{ce} }{t_{i} } .        
\end{equation} 

The free fall time in the original comoving system is given by (from Eqs. \eqref{ZEqnNum497055}, \eqref{ZEqnNum863096}, and \eqref{ZEqnNum156295})
\begin{equation} 
\label{ZEqnNum190681} 
\begin{split}
\frac{t_{c} }{s_{ce} }&=\frac{\sqrt{2} }{3\pi \lambda _{si} } \left(\exp \left[6\lambda _{c} \lambda _{si}^{2} \left(1+\sqrt{1+\frac{2}{\lambda _{si}^{2} } } \right)\right]-1\right)\\ 
&=\frac{t_{i} }{s_{ce} } \left(\exp \left[\frac{4\lambda _{c} }{3\pi ^{2} } \left(\frac{s_{ce} }{t_{i} } \right)^{2} \left(1+\sqrt{1+9\pi ^{2} \left(\frac{t_{i} }{s_{ce} } \right)^{2} } \right)\right]-1\right).
\end{split}
\end{equation} 
where two regimes can be clearly identified as, 
\begin{equation}
\begin{split}
&t_{c} =\frac{4}{\pi } \lambda _{c} s_{ce} \quad \textrm{for} \quad \lambda _{si}^{} \to 0\\ 
&\textrm{and}\\
&t_{c} =t_{i} \exp \left[\frac{8\lambda _{c} }{3\pi ^{2} } \left(\frac{s_{ce} }{t_{i} } \right)^{2} \right] \quad \textrm{for} \quad \lambda _{si}^{} \to \infty.  
\end{split}
\label{eq:58}
\end{equation}

\noindent The variation of the physical free fall time $t_{c}$ with $\lambda _{si}$ is also presented in Fig. \ref{fig:6} if correction factor $\lambda _{c} =1$ (the right axis). Conversely, Eq. \eqref{ZEqnNum190681} can be used to estimate the start time $t_{i} $ of a free fall if the free fall time $t_{c} $ is known. 

The free fall time $t_{c}$ increases if the same two-body system starts to collapse at an earlier time $t_{i} $. This is expected because the Hubble constant (damping) is greater at earlier time where larger resistance to the gravitational collapse is expected. Time $t_{c} $ approaches $s_{ce} $ when $s_{ce} $ is small (small separation or large mass) or when initial time $t_{i} $ approaches $t_{0} $. Clearly, larger $\lambda _{si}^{} $ (either greater separation between two body $r_{i} \to \infty $ or smaller total mass $M\to 0$ in Eq. \eqref{ZEqnNum252248}) or smaller $t_{i} $ (free fall starts at earlier physical time) will lead to a much larger free fall time than the exact free fall time in static background ($t_{c} \gg s_{ce} $).  

\subsection{TBCM model in the simplest form and perturbative \texorpdfstring{\\}{} solutions for equilibrium collapse}
\label{sec:3.4}
Next, the equation for $F\left(s\right)$ can be further simplified by introducing an amplitude function $F_{a} (\omega _{m} s)$. The original frequency function $F\left(s\right)$ can be decoupled into the product of the mean solution $F_{m} \left(s\right)$  (Eq. \eqref{ZEqnNum527908}) and an amplitude function $F_{a} (\omega _{m} s)$ as 

\begin{equation} 
\label{ZEqnNum688318} 
\begin{split}
F\left(s\right)&=F_{m} \left(s\right)F_{a} (\omega _{m} s)\\
&=\gamma _{s}^{-{1/\left(2+n\right)} } \left(\frac{r_{i} }{v_{i} } \right)^{{1/2} } \exp \left(-\frac{2-n}{2+n} \cdot \frac{H_{0} s}{4} \right)F_{a} (\omega _{m} s).
\end{split}
\end{equation} 
Substitution of Eq. \eqref{ZEqnNum688318} into the original Eq. \eqref{ZEqnNum502680} for $F\left(s\right)$ leads to a very simple equation for the amplitude function $F_{a} (x)$ with respect to a dimensionless variable $x=\omega _{m} \left(s\right)s$, 
\begin{equation}
\label{ZEqnNum129548} 
\frac{\partial ^{2} F_{a} (x)}{\partial x^{2} } =\underbrace{\frac{2n}{\left(2-n\right)^{2} } \frac{F_{a} (x)}{x^{2} } }_{1}-\underbrace{F_{a}^{n-1} (x)}_{2}+\underbrace{F_{a}^{-3} (x)}_{3},       
\end{equation} 
with initial conditions (using Eq. \eqref{ZEqnNum124395}),
\begin{equation}
\begin{split}
&F_{a} \left(x_{0} \right)=\gamma _{s}^{{1/\left(2+n\right)} } \quad \textrm{and} \quad \left. \frac{\partial F_{a} }{\partial x} \right|_{x=x_{0} } =\frac{\beta _{s} \gamma _{s}^{{-1/\left(2+n\right)} } }{2+n},\\ 
&\textrm{where} \quad x_{0} =\frac{2\gamma _{s}^{{2/\left(2+n\right)} } }{\beta _{s} } \frac{2+n}{2-n}.  
\end{split}
\label{ZEqnNum629408}
\end{equation}

\noindent The ratio $\gamma _{s} =\left({v_{ri} /v_{i} } \right)^{2} $ with $v_{ri} $ from Eq. \eqref{ZEqnNum443254}, $v_{i} $ is the initial speed, and the parameter ${\beta _{s} =4\lambda _{s} =H_{0} r_{i} /v_{i} } $ is introduced for convenience. Similarly, ${\beta _{si} =4\lambda _{si} =H_{0} r_{i} /v_{ri} } $ can be defined (Eq. \eqref{ZEqnNum252248}). 

Note that Eq. \eqref{ZEqnNum129548} is exact and is the simplest representation of original problem (Eq. \eqref{ZEqnNum365738}). Solution is fully determined by three dimensionless parameters, i.e. \textit{n}, $\beta _{s} $ and $\gamma _{s} $, a significant reduction from five parameters in original Eq. \eqref{ZEqnNum332742}.

For small \textit{x} or $n=-2$, term 1 (damping force) on the RHS of Eq. \eqref{ZEqnNum129548} is dominant and we have the exact solution of 
\begin{equation} 
\label{eq:62} 
F_{a} \left(x\to x_{0} \right)=\left(\frac{\beta _{s} \gamma _{s}^{-{1/2} } }{2} \frac{2-n}{2+n} x\right)^{{2/\left(2-n\right)} }  ,       
\end{equation} 
which is consistent with the transient solution in Eq. \eqref{ZEqnNum288261}. 

For large \textit{x}, term 2 (gravitational force) and term 3 (frequency force due to the angular momentum) are dominant. The trivial solution $F_{a} (x)=1$ can be easily identified for $\gamma _{s} =1$ and $\beta _{s} =0$ (i.e. static background without damping). If $n=-1$ and $\beta _{s} =0$, the original problem is reduced to the classical two-body gravitational problem in static background without damping. 

Here we focus on a more general case with a weak damping $\beta _{s} \to 0$ and $-2<n<0$ (large \textit{x} $x\ge x_{0} \gg 0$ from Eq. \eqref{ZEqnNum629408}), where the competition between terms 2 and 3 leads to an oscillatory solution vibrating around the mean value $F_{a} (x)=1$. It can be easily shown that if $F_{a} (x)<1$, we have $F_{a}^{-3} (x)>F_{a}^{n-1} (x)$, the positive curvature ${\partial ^{2} F_{a} (x)/\partial x^{2} } >0$ from Eq. \eqref{ZEqnNum129548} brings $F_{a} (x)$ back to $F_{a} (x)>1$; If $F_{a} (x)>1$, we have $F_{a}^{-3} (x)<F_{a}^{n-1} (x)$, curvature ${\partial ^{2} F_{a} (x)/\partial x^{2} } <0$ brings $F_{a} (x)$ back to the region $F_{a} (x)<1$; No oscillatory solution exists for short range force with $n\le -2$. 

We are especially interested in the oscillatory solutions with a weak damping ($\beta _{s} \to 0$), which is more relevant to the gravitational collapse in large-scale \textit{N}-body simulations. For weak damping, a harmonic function can be used to solve Eq. \eqref{ZEqnNum129548},
\begin{equation} 
\label{ZEqnNum472955} 
F_{a} (x)\approx A_{0} +A_{1} \sin \left[k_{s} \left(x-x_{0} \right)+A_{3} \right],        
\end{equation} 
where $A_{0} $ is the mean value, $A_{1} $ is the amplitude, $A_{3} $ is the phase angle, and $k_{s} $ is a dimensionless frequency. Substitution of Eq. \eqref{ZEqnNum472955} into Eq. \eqref{ZEqnNum129548}, the frequency $k_{s} $can be approximated by,
\begin{equation} 
\label{ZEqnNum918474} 
k_{s} \approx \left|\left(n-1\right)A_{0}^{n-2} +3A_{0}^{-4} \right|^{{1/2} } .          
\end{equation} 
To satisfy the boundary conditions (Eq. \eqref{ZEqnNum629408}), we have
\begin{equation}
\begin{split}
& A_{0} +A_{1} \sin \left(A_{3} \right)=\gamma _{s}^{{1/\left(2+n\right)} } \\ 
& \textrm{and} \\
& k_{s} A_{1} \cos \left(A_{3} \right)=\frac{\beta _{s} \gamma _{s}^{{-1/\left(2+n\right)} } }{\left(2+n\right)}.   
\end{split}
\label{ZEqnNum794685}
\end{equation}
\noindent Two limiting situations can be identified for weak damping ($\beta _{s} \to 0$):
\begin{enumerate}
\item\noindent Small initial velocity $v_{i} $ where $\gamma _{s} =\left({v_{ri} /v_{i} } \right)^{2} \gg 1$. This is the free fall collapse and free fall time is discussed in Section \ref{sec:3.3}. 
\item\noindent Large initial velocity $v_{i} $ where $\gamma _{s} =\left({v_{ri} /v_{i} } \right)^{2} \ll 1$. There exists a point in the trajectory with a vanishing kinetic energy and maximum potential (turning point). If this point is considered as the initial position, the trajectory after this point should be a free fall that is considered in 1). Therefore, both 1) and 2) will not likely lead to oscillatory motion. 
\item\noindent A more interesting case is the special case we discussed before, namely the initial velocity $v_{i} \approx v_{ri} $ and $\gamma _{s} =\left({v_{ri} /v_{i} } \right)^{2} \approx 1$ for initial system close to virial equilibrium. 
\end{enumerate}

Case (iii) leads to an equilibrium collapse with oscillatory solutions, as shown in Fig. \ref{fig:5}. For this case, $A_{0} =1\gg A_{1} $ (the fluctuation is small compared to the mean solution in Eq. \eqref{ZEqnNum472955}), $k_{s} =\sqrt{2+n} $ (from Eq. \eqref{ZEqnNum918474}), and $A_{3} =0$. Final perturbative solution for the amplitude function $F_{a} \left(x\right)$ (first order of $\beta _{s} $) reads
\begin{equation} 
\label{ZEqnNum372097} 
\begin{split}
F_{a} \left(x\right)&\approx 1+\frac{\beta _{s} }{\left(2+n\right)^{{3/2} } } \sin \left[\sqrt{2+n} \left(x-x_{0} \right)\right]\\
&=1+\frac{\beta _{s} }{\left(2+n\right)^{{3/2} } } \sin \left(\theta _{s} \left(x\right)\right),  
\end{split}
\end{equation} 
where the angle function $\theta _{s} $ is (with Eq. \eqref{ZEqnNum870001} for $\omega _{m} $ and Eq. \eqref{ZEqnNum758036} for mean radius $r_{m} $) 
\begin{equation} 
\label{ZEqnNum104758} 
\begin{split}
\theta _{s} \left(x\right)\equiv \theta _{s} \left(\omega _{m} s\right)&=\sqrt{2+n} \left(\omega _{m} s-x_{0} \right)\\
&=\frac{2\sqrt{2+n} }{\beta _{s} } \frac{2+n}{2-n} \left[\left(\frac{r_{m} }{r_{i} } \right)^{-{\left(2-n\right)/2} } -1\right], 
\end{split}
\end{equation} 
Or equivalently
\begin{equation}
\theta _{s} \left(s\right)=\frac{2\sqrt{2+n} }{\beta _{s} } \frac{2+n}{2-n} \left[\exp \left(\frac{2-n}{2+n}\frac{H_{0} s}{2} \right)-1\right]\approx \sqrt{2+n} \frac{sv_{i} }{r_{i} } 
\label{ZEqnNum426987}
\end{equation}
\noindent for $H_{0} s\ll1$, which approximates the angle swept by the displacement vector \textbf{\textit{r} }within time \textit{s} if $H_{0} s\ll 1$. Obviously solution \eqref{ZEqnNum372097} is valid only for $\beta _{s} <\left(2+n\right)^{{3/2} } $ such that the amplitude of oscillation is less than one in Eq. \eqref{ZEqnNum372097} for a positive amplitude function $F_{a} \left(x\right)$. This leads to the first critical value of $\beta_s$, just like the critical $c$ for over damped and under damped system in Eq. \eqref{eq:12}.

Solutions for $\beta _{s} =0$ and $n=-1$ are well known for two-body gravitational problem in static background (Kepler's law). Competition between gravity, damping, and angular momentum for $\beta _{s} \ne 0$ and $-2<n<0$ determines the free fall or equilibrium collapse for gravitational collapse in expanding background.

For a fixed mean radius $r_{m} \left(s\right)$, two-body systems with different initial separation $r_{i} $ can have different angle $\theta _{s} $ that is dependent on $r_{i} $ (Eq. \eqref{ZEqnNum104758}). Other relevant solutions can be found as,
\begin{equation} 
\label{ZEqnNum496296} 
\begin{split}
F\left(s\right)&=F_{m} \left(s\right)F_{a} \left(\omega \left(s\right)s\right)\\
&=\left(\frac{r_{i} }{v_{i} } \right)^{\frac{1}{2}} \exp \left(-\frac{2-n}{2+n}\frac{H_{0} s}{4} \right)\left\{1+\frac{\beta _{s} }{\left(2+n\right)^{{3/2} } } \sin \left(\theta _{s} \right)\right\}, 
\end{split}
\end{equation} 
\begin{equation} 
\label{ZEqnNum165747} 
\begin{split}
r\left(s\right)&=r_{m} \left(s\right)F_{a} \left(\omega _{m} \left(s\right)s\right)\\
&=r_{i} \exp \left(-\frac{H_{0} s}{2+n} \right)\left\{1+\frac{\beta _{s} }{\left(2+n\right)^{{3/2} } } \sin \left(\theta _{s} \right)\right\},
\end{split}
\end{equation} 
the frequency function $\omega \left(s\right)$ from Eq. \eqref{ZEqnNum553834}
\begin{equation} 
\label{eq:71} 
\omega \left(s\right)=\frac{1}{s} \int F_{m}^{-2} \left(s\right) F_{a}^{-2} \left(\omega _{m} \left(s\right)s\right)ds,        
\end{equation} 
and the time derivative of radius (or the radial velocity) 
\begin{equation} 
\label{ZEqnNum479123} 
\dot{r}=\frac{\partial r\left(s\right)}{\partial s} =\frac{H_{0} r_{i} }{\left(2+n\right)} \exp \left(-\frac{nH_{0} s}{2\left(2+n\right)} \right)\cos \left(\theta _{s} \right)-\frac{H_{0} r}{2+n} .     
\end{equation} 
Note that the first term on the RHS (right hand side) of Eq. \eqref{ZEqnNum479123} is from the time variation of angle $\theta _{s} $. This term becomes dominant over the second term with $r\to 0$ but can be averaged out for random $\theta _{s} $. This expression will be used to derive the stable clustering hypothesis (SCH) in Section \ref{sec:4.1} (Eq. \eqref{ZEqnNum390141}). For a given potential exponent \textit{n}, parameter $\beta _{s} $ controls both the amplitude and period of vibration (Eqs. \eqref{ZEqnNum426987} and \eqref{ZEqnNum496296}). 

The temporal evolution in transformed system with time scale \textit{s} can be equivalently transformed back to the evolution in original comoving system with physical time \textit{t} (Eq. \eqref{ZEqnNum185878}), where $s=t_{0} \ln \left({t/t_{i} } \right)$. Here $t_{i} $ (or $a_{i} $) is the initial time (or initial scale factor) and $t_{0} $ is the physical time of the present epoch. The exponential evolution with time \textit{s} is equivalent to a power-law evolution with physical time \textit{t},
\begin{equation} 
\label{ZEqnNum379959} 
\exp \left(\tau H_{0} s\right)\to \left({a/a_{i} } \right)^{\tau }.
\end{equation} 
Transforming back to comoving system, the mean separation $r_{m}$
\begin{equation}
\label{ZEqnNum966949} 
r_{m} \propto \exp \left(-{H_{0} s/\left(2+n\right)} \right)=\left({a/a_{i} } \right)^{-{1/\left(2+n\right)} }  
\end{equation} 
following a power-law can be obtained from Eq. \eqref{ZEqnNum758036} for the equilibrium range in Fig. \ref{fig:5}. Stable clustering hypothesis (SCH) refers to a comoving separation $r_{m} \propto a^{-1} $ or a fixed proper separation frozen in the physical time \textit{t}. Clearly, only the equilibrium collapse with $n=-1$ will lead to the stable clustering in expanding background (Eq. \eqref{ZEqnNum966949}). 
 
\subsection{Critical \texorpdfstring{$\beta _{s}$}{} for equilibrium collapse and critical density}
\label{sec:3.5}
For convenience, numerical constant $\beta _{s} =4\lambda _{s} ={H_{0} r_{i} /v_{i} }$ is introduced to quantify the competition between expanding background and gravity. Two critical values of $\beta _{s}$ can be identified from TBCM model and its solutions. 

Let's consider a two-body system that starts to collapse at an initial physical time $t_{i}$ with a corresponding Hubble constant $H_{i}$ and scale factor $a_{i}$. The evolution of angle function $\theta _{s}$ and separation $r$ in time scale \textit{s} (Eqs. \eqref{ZEqnNum426987} and \eqref{ZEqnNum165747}) can be equivalently transformed back to the evolution in physical time \textit{t}, where
\begin{equation} 
\label{ZEqnNum772779} 
\begin{split}
\theta _{s} \left(t\right)&=\frac{1}{2\lambda _{s} } \frac{\left(2+n\right)^{{3/2} } }{2-n} \left[\exp \left(\frac{2-n}{2+n} \cdot \frac{H_{0} s}{2} \right)-1\right]\\
&=\frac{2}{\beta _{s} } \frac{\left(2+n\right)^{{3/2} } }{2-n} \left(\left(\frac{t}{t_{i} } \right)^{\frac{2-n}{3\left(2+n\right)} } -1\right),
\end{split}
\end{equation} 
\begin{equation}
\label{ZEqnNum248861} 
r\left(t\right)=r_{i} \left(\frac{a_{i} }{a} \right)^{{1/\left(2+n\right)} } \left\{1+\frac{\beta _{s} }{\left(2+n\right)^{{3/2} } } \sin \left[\theta _{s} \left(t\right)\right]\right\} .      
\end{equation} 
Specifically, for $n=-1$,
\begin{equation}
\begin{split}
&r\left(t\right)=r_{i} \left(\frac{a_{i} }{a} \right)\left\{1+\beta _{s} \sin \left[\frac{2}{3\beta _{s} } \left(\frac{t}{t_{i} } -1\right)\right]\right\} \\
&\textrm{and} \\
&\theta _{s} \left(t\right)=\frac{2}{3\beta _{s} } \left(\frac{t}{t_{i} } -1\right).   
\end{split}
\label{eq:77}
\end{equation}

The first critical value of $\beta _{s} $ can be identified for the existence of an equilibrium range (under damped in Fig. \ref{fig:5}) from Eq. \eqref{ZEqnNum248861},
\begin{equation} 
\label{ZEqnNum552089} 
\beta _{s1} =\frac{H_{0} r_{ic} }{v_{ic} } =\left(2+n\right)^{{3/2} } ,          
\end{equation} 
which leads to a critical initial separation $r_{ic} $ or initial velocity $v_{ic} $ when combined with Eq. \eqref{ZEqnNum443254},
\begin{equation}
\begin{split}
&r_{ic} =\frac{1}{2} \left[\frac{-nG_{n} \left(m_{1} +m_{2} \right)}{H_{0}^{2} \left(2+n\right)^{-3} } \right]^{{1/\left(2-n\right)} }\\
&\textrm{and}\\
&v_{ic} =\frac{1}{2} \frac{\left[-nG_{n} \left(m_{1} +m_{2} \right)\right]^{{1/\left(2-n\right)} } }{\left[H_{0}^{} \left(2+n\right)^{-{3/2} } \right]^{{n/\left(2-n\right)}}}.   
\end{split}
\label{ZEqnNum562446}
\end{equation}

\noindent Both $r_{ic} $ and $v_{ic} $ are only dependent on the system mass, the damping $H_{0}^{} $ and the potential exponent \textit{n}. For $m_{1} =m_{2} =2.27\times 10^{11} {M_{sun} /h} $, $r_{ic} =0.29{Mpc/h} $ and $v_{ic} =29{km/s} $. This is the maximum separation and the corresponding velocity for the existence of an equilibrium two-body collapse.

For systems that evolve from initial virial equilibrium (the special case $\gamma _{s} =\left({v_{ri} /v_{i} } \right)^{2} =1$), the equilibrium collapse exists only if $\beta _{s} \le \beta _{s1} =\left(2+n\right)^{{3/2} } $ (or $r_{i} <r_{ic} $ or $v_{i} =v_{ri} >v_{ic} $), where gravity is sufficiently large to balance expanding background in order to form the equilibrium collapse. For $\beta _{s} >\beta _{s1} =\left(2+n\right)^{{3/2} } $, gravity is too weak to establish an equilibrium collapse and system is over damped. 

Next, the second critical value of $\beta_s$ can be obtained by considering a continuous growth of a halo starting from $t=0$ to $t=t_{0}$ with an infinitesimal lifetime and extremely fast mass accretion. This halo is formed by continuously growing via a sequence of two-body collapse (merging) events with single mergers with an infinitesimal waiting time. This is a good approximation for large halos as the lifetime of halo $\tau_g \propto m_h^{-2/3}$ \citep[see][Eq. (45)]{Xu:2021-Inverse-mass-cascade-mass-function}. Therefore, halos with an infinitesimal lifetime have no time to relax and should always follow the mean solutions of radius $r_m$ and frequency $F_m$. The phase angle should vanish in Eq. \eqref{ZEqnNum248861} such that
\begin{equation}
\sin \left[\theta _{s} \left(t\right)\right]=0 \quad \textrm{for any} \quad t\in \left[0,t_{0} \right],       
\label{ZEqnNum657387}
\end{equation}
where $r\left(t\right)=r_{m} $ in Eqs. \eqref{ZEqnNum165747} and \eqref{ZEqnNum248861}  from $t=0$ to $t=t_{0} $ without oscillation. Let us assume the first merging event occurs at time $t_{i} $. With $t_{i} \to 0$, we can safely assume that $k={t/t_{i} } $ is an integer. From Eqs. \eqref{ZEqnNum772779} and \eqref{ZEqnNum657387}, 
\begin{equation} 
\label{ZEqnNum193909} 
\sin \left[\theta _{s} \left(t\right)\right]=\sin \left[\frac{2}{\beta _{s} } \frac{\left(2+n\right)^{{3/2} } }{2-n} \left(k^{\frac{2-n}{3\left(2+n\right)} } -1\right)\right]=0.      
\end{equation} 
The second critical value of $\beta _{s} $ can be identified from Eq. \eqref{ZEqnNum193909} for any arbitrary integer $k$, 
\begin{equation} 
\label{ZEqnNum937360} 
\beta _{s2} =\frac{\left(2+n\right)^{{3/2} } }{\left(2-n\right)\pi }.
\end{equation} 
Note that there exists a constant value of $\beta _{s2} $ satisfying Eq. \eqref{ZEqnNum193909} for any integer $k$ if and only if
\begin{equation}
\begin{split}
&m=\frac{2-n}{3\left(2+n\right)} \quad \textrm{and} \quad n=\frac{2-6m}{1+3m}=-1,-\frac{10}{7},-\frac{8}{5}...-2 \\ &\textrm{with} \quad m=1,2,...\infty, \textrm{and}\\
&\beta _{s2}=\frac{2}{3\pi m\sqrt{1+3m} }=\frac{1}{3\pi},\textrm{ }\frac{1}{3\pi\sqrt{7}}...0. 
\end{split}
\label{eq:83}
\end{equation}

The parameter $\beta _{s} $ for halos with infinitesimal waiting time should always satisfy $\beta _{s} =\beta _{s2} $. The time derivative of angle $\theta _{s} $ (angular speed) can be easily obtained from Eq. \eqref{ZEqnNum772779},
\begin{equation} 
\label{eq:84} 
\left. \frac{d\theta _{s} }{dt} \right|_{t=t_{i} } =\frac{2\sqrt{2+n} }{3\beta _{s2} t} =\frac{\left(2-n\right)}{\left(2+n\right)} \frac{2\pi }{3t} ,        
\end{equation} 
from which we can find the period $T_{s} $ for large halos formed at any instant time \textit{t,} 
\begin{equation} 
\label{ZEqnNum256922} 
T_{s} =\frac{2\pi }{{d\theta _{s} /dt} } =\frac{3\left(2+n\right)}{\left(2-n\right)} t=\frac{t}{m} .         
\end{equation} 
Specifically, for large halos with an infinitesimal lifetime and $n=-1$, halos formed at any instant \textit{t} from a two-body collapse with a single merger has a period of $T_{s} =t$ (the orbital period of outer region of halos should be comparable to the current physical time \textit{t}). 

Two numerical constants $\alpha _{s} $ (in Eq. \eqref{ZEqnNum577609}) and $\beta _{s} ={H_{0} r_{i} /v_{i} } $ are closely related to the density ratio of two-body system to the background. The critical density ratio $\Delta_c=18\pi^2$ (usually derived from spherical collapse model) can be simply derived by our two-body collapse model (TBCM) as follows: 

Let's consider halos start equilibrium collapse at physical time $t_{i} $ with a corresponding Hubble constant $H_{i} $ ($H_{0}^{2} =H_{i}^{2} a_{i}^{3} $) and scale factor $a_{i} $. The two numerical constants are defined as 
\begin{equation} 
\label{ZEqnNum149058} 
\begin{split}
&\beta _{s} =\frac{H_{0} r_{i} }{v_{i} } =\frac{H_{0} r_{i} }{u_{i} a_{i}^{{1/2} } } =\frac{H_{i} r_{yi} }{u_{i} }\\
&\textrm{and}\\
&\alpha _{s} =\frac{v_{i}^{2} r_{i}^{-n} }{G_{n} m_{h} } =\frac{u_{i}^{2} a_{i} r_{i}^{-n} }{G_{n} m_{h} } =\frac{u_{i}^{2} r_{yi}^{-n} a_{i}^{n+1} }{G_{n} m_{h} }.
\end{split}
\end{equation} 

Here $\alpha _{s} $ is the virial constant from Eq. \eqref{ZEqnNum577609} , $u_{i} =v_{i} a_{i}^{{-1/2} } $ is the peculiar velocity at time $t_{i} $, $v_{i} $ is the velocity in transformed system with time scale \textit{s}, and $r_{yi} =a_{i} r_{i} $ is the separation in physical coordinate at time $t_{i} $. Large halos with infinitesimal lifetime are synchronized. All halos are generated at the same time \textit{t} and both constants $\alpha _{s} $ and $\beta _{s} $ should approach constant values (Eqs. \eqref{ZEqnNum937360} and \eqref{ZEqnNum577609}), i.e a direct delta distribution. For small halos with a finite lifetime, there can be a distribution of values of $\alpha _{s} $ and $\beta _{s} $ since small halos are generated at different initial time $t_{i} $ can co-exist at the same time \textit{t}. The mass dependence of both parameters is presented in a separate paper \citep[see][Fig. 2]{Xu:2022-The-evolution-of-energy--momen}.

Note that $H^{2} ={8\pi G\bar{\rho }_{y} \left(t\right)/3} $, where $\bar{\rho }_{y} \left(t\right)$ is the physical density of background, Eq. \eqref{ZEqnNum149058} can be used to derive a ratio $\Delta_c$ of the physical density of halos to the background density at time $t_{i} $, 
\begin{equation} 
\label{ZEqnNum776754} 
\Delta_c =\frac{\rho _{s} \left(t_{i} \right)}{a_{i}^{3} \bar{\rho }_{y} \left(t_{i} \right)} =\frac{\rho _{y} \left(t_{i} \right)}{\bar{\rho }_{y} \left(t_{i} \right)} =\frac{1}{4\alpha _{s} \beta _{s}^{2} } \frac{G}{G_{n} r_{i}^{1+n} } .       
\end{equation} 
Here $\rho _{y} $ is the mean physical density of halo. Comoving density $\rho _{s} $ of the two-body system is 
\begin{equation} 
\label{eq:88} 
\rho _{s} \left(t_{i} \right)=\frac{M}{{4\pi \left(2r_{i} \right)^{3} /3} } ,          
\end{equation} 
where $M=m_{1} +m_{2} $ and the halo size is $r_{h} =2r_{i} $ because of $m_{1} \ll m_{2} $ (large halo merges with a single merger where the mass of a single merger is much smaller) and $\mu =2$ (in Eq. \eqref{ZEqnNum365738}). 

The critical density can be computed based on two critical values $\beta _{s1} $ and $\beta _{s2} $. For gravitational collapse of a two-body system with $n=-1$, $G=G_{n} $, $\alpha _{s} ={-n/2^{2-n} } ={1/8} $ (Eq. \eqref{ZEqnNum577609}), and $\beta _{s} <\beta _{s1} =1$, only the system with a physical density $\rho _{y} \left(t_{i} \right)>2\bar{\rho }_{y} \left(t_{i} \right)$ (from Eq. \eqref{ZEqnNum776754}) will lead to an equilibrium collapse. Systems with a physical density $\rho _{y} \left(t_{i} \right)<2\bar{\rho }_{y} \left(t_{i} \right)$ will have a free fall collapse that can be completed in a much short period (Fig. \ref{fig:5}). 

For $n=-1$, the density ratio $\Delta $ of large halos with an infinitesimal lifetime and $\beta _{s} =\beta _{s2} $ is (from Eqs. \eqref{ZEqnNum776754}, \eqref{ZEqnNum937360}, and $\alpha _{s} ={-n/2^{2-n} } $),
\begin{equation}
\Delta_c =\frac{\rho _{y} \left(t_{i} \right)}{\bar{\rho }_{y} \left(t_{i} \right)} =\frac{1}{4\alpha _{s} \beta _{s2}^{2} } =-\frac{2^{-n} \left(2-n\right)^{2} }{n\left(2+n\right)^{3} } \pi ^{2} =18\pi ^{2} \textrm{ for } \textrm{n=-1}.    
\label{ZEqnNum979432}
\end{equation}

\noindent Surprisingly, this critical density ratio is consistent with the prediction from spherical collapse model (SCM) and reveals deep connections between TBCM and SCM models. More discussion will be presented in Sections \ref{sec:4.3}.  

\subsection{Solutions for energy, virial quantity, and angular momentum}
\label{sec:3.6}
We can demonstrate that the specific kinetic and potential energies (per unit mass) for two-body collapsing system are evolving exponentially in the time scale \textit{s} (or equivalently a power-law with respect to \textit{a} in the original comoving system using Eq. \eqref{ZEqnNum379959}). The specific kinetic energy reads (from Eqs. \eqref{ZEqnNum435078}, \eqref{ZEqnNum542855}, and \eqref{ZEqnNum553834})
\begin{equation}
\label{ZEqnNum369546} 
\begin{split}
K_{s}&=\frac{1}{\left(m_{1} +m_{2} \right)} \left[\frac{1}{2} m_{1} \left(v_{x1}^{2} +v_{y1}^{2} \right)+\frac{1}{2} m_{2} \left(v_{x2}^{2} +v_{y2}^{2} \right)\right]\\
&=\frac{2m_{1} m_{2} }{\left(m_{1} +m_{2} \right)^{2} } \left[\underbrace{\dot{r}^{2} }_{1}+\underbrace{r^{2} F\left(s\right)^{-4} }_{2}\right].  
\end{split}
\end{equation} 

The first term (term1) on the RHS represents the contribution from the radial motion that is small when compared to the second term. The ratio between two terms on the RHS of Eq. \eqref{ZEqnNum369546} can be obtained from Eqs. \eqref{ZEqnNum165887} and \eqref{ZEqnNum496296},
\begin{equation} 
\label{eq:91} 
\left[\frac{\dot{r}}{rF\left(s\right)^{-2} } \right]^{2} =F^{4} \left(\frac{\partial \ln F}{\partial s} -\frac{H_{0} }{4} \right)^{2} \approx \frac{\beta _{s}^{2} }{\left(2+n\right)^{2} } \exp \left(-\frac{2-n}{2+n} \cdot H_{0} s\right).    
\end{equation} 
For small $\beta _{s} $, this ratio is exponentially decaying with time \textit{s} and proportional to $\beta _{s}^{2} $ (second order). By neglecting the high order term (term 1) and using Eqs. \eqref{ZEqnNum496296} for $F\left(s\right)$ and Eq. \eqref{ZEqnNum165747} for $r\left(s\right)$, the final expression of the specific kinetic energy reads
\begin{equation} 
\label{ZEqnNum521799} 
K_{s} \approx \frac{2m_{1} m_{2} v_{i}^{2} }{\left(m_{1} +m_{2} \right)^{2} } \exp \left(\frac{-nH_{0} s}{2+n} \right)\left[1-\frac{2\beta _{s} }{\left(2+n\right)^{{3/2} } } \sin \theta _{s} \right].      
\end{equation} 

Similarly, the specific potential energy reads (with the expression of \textit{r} from Eq. \eqref{ZEqnNum165747} and $v_{i}^{2} $ from Eq. \eqref{ZEqnNum577609}), 
\begin{equation} 
\label{ZEqnNum978254} 
\begin{split}
P_{s}&=-\frac{G_{n} m_{1} m_{2} }{\left(2r\right)^{-n} } \frac{1}{\left(m_{1} +m_{2} \right)} \\
&\approx \frac{2m_{1} m_{2} v_{i}^{2} }{\left(m_{1} +m_{2} \right)^{2} } \exp \left(\frac{-nH_{0} s}{2+n} \right)\left[\frac{2}{n} +\frac{2\beta _{s} }{\left(2+n\right)^{{3/2} } } \sin \theta _{s} \right].
\end{split}
\end{equation} 

Total energy for two-body system and each individual mass are
\begin{equation} 
\label{ZEqnNum387242} 
\begin{split}
E_{s} =K_{s} +P_{s} &=\frac{\left(2+{4/n} \right)m_{1} m_{2}^{} v_{i}^{2} }{\left(m_{1} +m_{2} \right)^{2} } \exp \left(\frac{-nH_{0} s}{2+n} \right)\\
&=\frac{-\left(2+n\right)m_{1} m_{2}^{} }{\left(m_{1} +m_{2} \right)} \frac{G_{n} r_{i} }{\left(2r_{i} \right)^{1-n} } \exp \left(\frac{-nH_{0} s}{2+n} \right),
\end{split}
\end{equation} 
\begin{equation} 
\label{ZEqnNum588856} 
\begin{split}
E_{s1}&=\left(\underbrace{\frac{-nm_{2}^{2} }{m_{1} +m_{2} } }_{kinetic}\underbrace{-m_{2} }_{potential}\right)\frac{\left(-2\right)v_{i}^{2} }{n\left(m_{1} +m_{2} \right)} \exp \left(\frac{-nH_{0} s}{2+n} \right)\\
&=\left(\underbrace{\frac{-nm_{2}^{2} }{m_{1} +m_{2} } }_{kinetic}\underbrace{-m_{2} }_{potential}\right)\frac{G_{n} r_{i} }{\left(2r_{i} \right)^{1-n} } \exp \left(\frac{-nH_{0} s}{2+n} \right),  
\end{split}
\end{equation} 
\begin{equation} 
\label{ZEqnNum634242} 
\begin{split}
E_{s2}&=\left(\frac{-nm_{1}^{2} }{m_{1} +m_{2} } -m_{1} \right)\frac{\left(-2\right)v_{i}^{2} }{n\left(m_{1} +m_{2} \right)} \exp \left(\frac{-nH_{0} s}{2+n} \right)\\
&=\left(\frac{-nm_{1}^{2} }{m_{1} +m_{2} } -m_{1} \right)\frac{G_{n} r_{i} }{\left(2r_{i} \right)^{1-n} } \exp \left(\frac{-nH_{0} s}{2+n} \right),
\end{split}
\end{equation} 
respectively. As shown in Fig. \ref{fig:3}, both $K_{s} $ and $P_{s} $ vibrate about their mean solutions with an amplitude proportional to the parameter $\beta _{s} $ to the first order. The specific energy $E_{s} $ does not vibrate due to the cancellation of first order perturbation in $K_{s} $ and $P_{s} $ (Eq. \eqref{ZEqnNum387242}). 

By considering an ensemble of many two-body systems with randomly distributed angles $\theta _{s} $, the ensemble average of kinetic and potential energies of these two-body systems are
\begin{equation} 
\label{ZEqnNum187238} 
\begin{split}
&\left\langle K_{s} \right\rangle =\frac{2m_{1} m_{2} v_{i}^{2} }{\left(m_{1} +m_{2} \right)^{2} } \exp \left(\frac{-nH_{0} s}{2+n} \right)\\ 
&\textrm{and}\\ 
&\left\langle P_{s} \right\rangle =\frac{4m_{1} m_{2} v_{i}^{2} }{n\left(m_{1} +m_{2} \right)^{2} } \exp \left(\frac{-nH_{0} s}{2+n} \right),  
\end{split}
\end{equation} 
where first order perturbations are averaged out. The average kinetic and potential energy satisfy the virial equilibrium, where $2\left\langle K_{s} \right\rangle -n\left\langle P_{s} \right\rangle =0$ in the equilibrium range. 

The system spends most time in the equilibrium range with an exponential evolution of energy in time scale \textit{s} (from Eqs. \eqref{ZEqnNum521799}, \eqref{ZEqnNum978254} and \eqref{ZEqnNum187238}). Equivalently, energy follows a power-law evolution in physical time \textit{t}, i.e.\textit{ }$\left\langle K_{s} \right\rangle \propto t$ and $\left\langle P_{s} \right\rangle \propto t$ for n=-1, that will provide some clues for the energy evolution in large scale N-body system, as discussed in a separate paper \citep[see][Fig. 1a]{Xu:2022-The-evolution-of-energy--momen}. More importantly, this also hints a constant rate of energy cascade $\epsilon_u$ in dark matter flow \citep{Xu:2022-The-evolution-of-energy--momen,Xu:2021-Inverse-and-direct-cascade-of-}.

More interestingly for \textit{n}=-1, the evolution of specific energy of two individual particles (Eqs. \eqref{ZEqnNum588856} and \eqref{ZEqnNum634242}) is the same for both particles regardless of their masses, where $E_{s} =E_{s1} =E_{s2} $. For a two-body system with unequal mass $m_{1} \ne m_{2} $, the specific energy is independent of particle mass ($E_{s1} =E_{s2} $), while the total energy is proportional to particle mass. The energy equipartition does not apply for the two-body system in equilibrium range, where there is no energy transfer between two particles. The energy evolution does not depend on individual mass, which seems consistent with the concept of violent relaxation. More discussion is presented in Section \ref{sec:4.2}.   

The temporal evolution of the specific virial quantity $G_{s} $ (mass averaged radial velocity moment) can be found from Eqs. \eqref{ZEqnNum186239} and \eqref{ZEqnNum988334}, where $G_{s} $ is defined as,
\begin{equation} 
\label{ZEqnNum259472} 
\begin{split}
G_{s} =\frac{\sum m_{i} \boldsymbol{\mathrm{x}}_{i} \cdot \boldsymbol{\mathrm{v}}_{i}  }{\Sigma m_{i} } &=\frac{m_{1} \boldsymbol{\mathrm{x}}_{1} \cdot \boldsymbol{\mathrm{v}}_{1} +m_{2} \boldsymbol{\mathrm{x}}_{2} \cdot \boldsymbol{\mathrm{v}}_{2} }{m_{1} +m_{2} }\\
&=\frac{4m_{1} m_{2} }{\left(m_{1} +m_{2} \right)^{2} } \boldsymbol{\mathrm{r}}\cdot \boldsymbol{\mathrm{v}}=\frac{4m_{1} m_{2} }{\left(m_{1} +m_{2} \right)^{2} } \dot{r}r.  
\end{split}
\end{equation} 
Using Eqs. \eqref{ZEqnNum165747} and \eqref{ZEqnNum479123} for radius $r$ and $\dot{r}$, the specific virial quantity $G_{s} $ can be written as,
\begin{equation} 
\label{ZEqnNum463637} 
\begin{split}
G_{s} =\frac{4m_{1} m_{2} }{\left(m_{1} +m_{2} \right)^{2} } &\left\{\frac{H_{0} r_{i}^{2} }{\left(2+n\right)} \exp \left(-\frac{H_{0} s}{2} \right)\left(\cos \theta _{s} +\frac{\beta _{s} }{2\left(2+n\right)^{{3/2} } } \sin 2\theta _{s} \right)\right.\\ &\left.-\frac{H_{0} r_{i}^{2} }{2+n} \exp \left(-\frac{2H_{0} s}{2+n} \right)\left(1+\frac{\beta _{s} }{\left(2+n\right)^{{3/2} } } \sin \theta _{s} \right)^{2} \right\}.   
\end{split}
\end{equation} 
Similarly, the specific angular momentum of the two-body system can be obtained as,
\begin{equation} 
\label{ZEqnNum129930} 
\begin{split}
\boldsymbol{\mathrm{H}}_{s}&=\frac{m_{1} \boldsymbol{\mathrm{x}}_{1} \times \boldsymbol{\mathrm{v}}_{1} +m_{2} \boldsymbol{\mathrm{x}}_{2} \times \boldsymbol{\mathrm{v}}_{2} }{m_{1} +m_{2} } =\frac{4m_{1} m_{2} }{\left(m_{1} +m_{2} \right)^{2} } \boldsymbol{\mathrm{r}}\times \boldsymbol{\mathrm{v}}\\
&=\frac{4m_{1} m_{2} }{\left(m_{1} +m_{2} \right)^{2} } r^{2} F^{-2} \left(s\right)\hat{\boldsymbol{\mathrm{z}}}=\frac{4m_{1} m_{2} v_{i} r_{i}^{} }{\left(m_{1} +m_{2} \right)^{2} } \exp \left(-\frac{1}{2} H_{0} s\right)\hat{\boldsymbol{\mathrm{z}}}.
\end{split}
\end{equation} 
The angular momentum $\boldsymbol{\mathrm{H}}_{s} $ decays exponentially at a rate of ${H_{0} /2} $ that is independent of the potential exponent \textit{n}. 

\subsection{Two-body angular velocity \texorpdfstring{$\omega _{t}$}{}, angle of incidence \texorpdfstring{$\theta _{\boldsymbol{\mathrm{vr}}},$\\}{} and halo kinetic energy}
\label{sec:3.7}
The two-body collapse model (TBCM) and its solutions are presented. The two critical density ratios are identified. Rich information contained in TBCM model can be used to provide more insights into the structure formation and energy evolution. This section presents several additional applications of TBCM.

The first example is about the two-body angular velocity $\omega _{s} $ that can be found from the kinetic energy solution with
\begin{equation} 
\label{ZEqnNum185223} 
\frac{1}{2} \left(m_{1} \mu ^{2} +m_{2} \left(2-\mu \right)^{2} \right)\omega _{s}^{2} r^{2} =\left(m_{1} +m_{2} \right)K_{s} .       
\end{equation} 
With $r$ from Eq. \eqref{ZEqnNum165747}, $\mu ={2m_{2} /\left(m_{1} +m_{2} \right)} $ from Eq. \eqref{ZEqnNum365738}, and kinetic energy from Eq. \eqref{ZEqnNum521799}, the angular velocity $\omega _{s} $ in transformed system is obtained from Eq. \eqref{ZEqnNum185223},
\begin{equation} 
\label{eq:102} 
\omega _{s} \approx \frac{v_{i}^{} }{r_{i} } \exp \left[\frac{2-n}{2\left(2+n\right)} H_{0} s\right].         
\end{equation} 
For $n=-1$, angular velocity $\omega _{t} $ in original comoving system (with $\gamma _{s} =1$ and $r_{m} $ in Eq. \eqref{ZEqnNum758036} and Eq. \eqref{ZEqnNum379959} for transformation) is, 
\begin{equation} 
\label{ZEqnNum762297} 
\omega _{t} =\omega _{s} \frac{ds}{dt} =\omega _{s} a^{{-3/2} } =\frac{r_{i}^{{3/2} } }{\beta _{s} } Hr_{m}^{{-3/2} } ,      \end{equation} 
where the two-body angular velocity $\omega _{t} \sim Hr_{m}^{-{3/2} } $ is inversely proportional to the mean separation $r_{m}$ and is proportional to the Hubble parameter. This can be confirmed by N-body simulation in separate papers \citep[see][Fig. 15]{Xu:2021-Inverse-and-direct-cascade-of-} \citep[also see][Fig. 3]{Xu:2022-The-mean-flow--velocity-disper}. 

The second example is about the angle of incidence $\theta _{\boldsymbol{\mathrm{vr}}}$, i.e. the angle between particle velocity and the vector of separation. The virial quantity $G_{s} $ (Eqs. \eqref{ZEqnNum259472} and \eqref{ZEqnNum463637}) represents the relative motion of two particles in the radial direction, while $\boldsymbol{\mathrm{H}}_{s} $ (Eq. \eqref{ZEqnNum129930}) stands for the relative motion in the tangential direction. Terms involving $\theta _{s} $ in Eq. \eqref{ZEqnNum463637} can be averaged out when averaging over many two-body systems with random angle $\theta _{s} $. The angle $\theta _{\boldsymbol{\mathrm{vr}}} $ between the displacement vector \textbf{\textit{r}} and its velocity vector \textbf{\textit{v} }can be computed using Eqs. \eqref{ZEqnNum463637}, \eqref{ZEqnNum129930}, Eq. \eqref{ZEqnNum165747} for \textit{r,} and the transformation between time scales $s=t_{0} \ln \left({t/t_{i} } \right)$,
\begin{equation} 
\label{ZEqnNum994889} 
\begin{split}
\cot \left(\theta _{\boldsymbol{\mathrm{vr}}} \right)&=-\frac{v_{r} }{v_{cir} } =\frac{\boldsymbol{\mathrm{r}}\cdot \boldsymbol{\mathrm{v}}}{\left|\boldsymbol{\mathrm{r}}\times \boldsymbol{\mathrm{v}}\right|} =\frac{G_{s} }{\left|\boldsymbol{\mathrm{H}}_{s} \right|}\\
&\approx -\frac{H_{0} r_{i} }{v_{i} \left(2+n\right)} \exp \left(\frac{n-2}{2\left(n+2\right)} H_{0} s\right)=-\frac{\beta _{s} }{\left(2+n\right)} \left(\frac{a}{a_{i} } \right)^{\frac{n-2}{2\left(n+2\right)} }, 
\end{split}
\end{equation} 
where $v_{r} $ is the radial velocity and $v_{cir} $ is the circular velocity. For $\beta _{s} =\beta _{s2} $, i.e. halos with an infinitesimal lifetime, the angle $\theta _{\boldsymbol{\mathrm{vr}}} $ is slightly $>{\pi /2} $ (i.e. $\cot \left(\theta _{\boldsymbol{\mathrm{vr}}} \right)\approx \cos \left(\theta _{\boldsymbol{\mathrm{vr}}} \right)$) due to the gravitational interaction. Equation \eqref{ZEqnNum994889} predicts that for two-body system, the angle $\theta _{\boldsymbol{\mathrm{vr}}} $ between the pairwise velocity $\Delta \boldsymbol{\mathrm{u}}=\boldsymbol{\mathrm{u}}_{1} -\boldsymbol{\mathrm{u}}_{2} $ and their separation vector $\Delta \boldsymbol{\mathrm{r}}=\boldsymbol{\mathrm{r}}_{1} -\boldsymbol{\mathrm{r}}_{2} $ satisfies
\begin{equation} 
\label{eq:105} 
\cot \left(\theta _{\boldsymbol{\mathrm{vr}}} \right)=-\beta _{s} \left(\frac{a}{a_{i} } \right)^{{-3/2} } =-\beta _{s} \left(\frac{r_{m} }{r_{i} } \right)^{{3/2} } ,        
\end{equation} 
where $r_{m} $ is the mean separation and $\theta _{\boldsymbol{\mathrm{vr}}} \to {\pi /2} $ with time \textit{t} or \textit{a}. For two-body collapse between large halos with an infinitesimal lifetime and a single merger, $\beta _{s} =\beta _{s2} $ and $\cot \left(\theta _{\boldsymbol{\mathrm{vr}}} \right)=-{1/\left(3\pi \right)} $ such that the angle between the velocity $\boldsymbol{\mathrm{v}}$ of that single merger at halo surface and its position vector $\boldsymbol{\mathrm{r}}$ from halo center should be $\theta _{\boldsymbol{\mathrm{vr}}} \approx 96.06^{o} $, which is consistent with the result for halos with an isothermal density profile \citep[see][Eq. (31)]{Xu:2021-Inverse-mass-cascade-halo-density}. The ratio of radial velocity of single merger to its circular velocity is always a constant ${v_{r} /v_{cir} } =\beta _{s2} ={1/3\pi}$. This angle represents the angle of incidence when single merger merges with halos. It is relevant to the interpretation of critical acceleration $a_0$ for MOND theory (modified Newtonian dynamics) \citep[see][Fig. 8 and Eq. (12)]{Xu:2022-The-origin-of-MOND-acceleratio}.

Finally, halo with an isothermal density profile can be a direct result of infinitesimal waiting time such that the radial flow vanishes \citep[see][Section 3.3]{Xu:2021-Inverse-mass-cascade-halo-density}. This point can also be demonstrated by the application of TBCM model to derive the halo energy. The last example is to derive the kinetic and potential energy for large halos with an infinitesimal lifetime. 

Let's consider a halo of mass \textit{M} with a specific peculiar kinetic energy $K_{h} $ that is continuously growing via elementary two-body merging with a single merger of mass \textit{dM} during an infinitesimal time \textit{dt}. Since $dt\approx 0$, the change of total kinetic energy of two-body system in transformed system will be $\left(M+dM\right)\cdot K_{s} \left(s=0\right)$ with $K_{s} $ from Eq. \eqref{ZEqnNum521799}. The incremental change of the specific peculiar kinetic energy in comoving system from a single merging event is
\begin{equation} 
\label{ZEqnNum198910} 
dK_{h} =\frac{K_{s} \left(s=0\right)}{a} =\frac{2M}{M^{2} } dM\frac{v_{i}^{2} }{a} =\frac{2M}{M^{2} } dM\cdot \alpha _{s} \frac{G_{n} M}{ar_{i}^{-n} } ,      
\end{equation} 
where the last equality is from the fact that $v_{i}^{2} =\alpha _{s} {G_{n} M/r_{i} {}^{-n} } $ (Eq. \eqref{ZEqnNum577609}) and $r_{i} {}^{} $ is the comoving length of the displacement vector. Since $dt\approx 0$, the merging is instantaneous. The halo mass $M\propto r_{i}^{3} $ and the halo kinetic energy $K_{h} \propto {G_{n} M/r_{i} {}^{-n} \propto M} ^{1+{n/3} } $ for halos of different mass at the same redshift (virial theorem). From Eq. \eqref{ZEqnNum198910}
\begin{equation} 
\label{eq:107} 
\frac{d\ln K_{h} }{d\ln M} =\frac{2\alpha _{s} G_{n} M}{K_{h} ar_{i}^{-n} } =1+\frac{n}{3} .         
\end{equation} 
The final expressions for halo kinetic and potential energy are: (using the virial theorem $2K_{h} -n_{e}^{} P_{h} =0$, where $n_{e} \approx -1.3$ is an effective exponent due to halo surface energy and $n_{e} \approx -1.5$ for halos with an isothermal density \citep[see][Eq. (96)]{Xu:2021-Inverse-mass-cascade-halo-density}
\begin{equation}
K_{h} =\frac{6\alpha _{s} }{3+n} \frac{G_{n} M}{ar_{i}^{-n} } \quad  \textrm{and} \quad P_{h} =\frac{12}{\left(3+n\right)} \frac{\alpha _{s} }{n_{e} } \frac{G_{n} M}{ar_{i}^{-n} }.     
\label{ZEqnNum235138}
\end{equation}

\noindent The one-dimensional velocity dispersion
\begin{equation} 
\label{ZEqnNum947544} 
\sigma _{v}^{2} =\frac{2}{3} K_{h} =\frac{4\alpha _{s} }{3+n} \frac{G_{n} M}{ar_{i}^{-n} } ,         
\end{equation} 
with $\alpha _{s} ={-n/2^{2-n} } $ defined in Eq. \eqref{ZEqnNum577609}. Specifically, for $n=-1$,
\begin{equation} 
\label{ZEqnNum157977} 
K_{h} =\frac{3}{8} \frac{GM}{ar_{i}^{} } =\frac{3}{8} \frac{GM}{r_{yi}^{} } =\frac{3}{4} \frac{GM}{r_{h} } , P_{h} =\frac{3}{4} \frac{GM}{n_{e} ar_{i}^{} } =\frac{3}{2n_{e} } \frac{GM}{r_{h} } ,      
\end{equation} 
and halo virial dispersion 
\begin{equation} 
\label{ZEqnNum786168} 
\sigma _{v}^{2} =\frac{2}{3} K_{h} =2\frac{v_{i}^{2} }{a} =2u_{i}^{2} =\frac{1}{4} \frac{GM}{ar_{i}^{} } =\frac{GM}{2r_{h} } ,       
\end{equation} 
where $r_{yi}^{} =ar_{i} ={r_{h} /2} $ is the length of displacement vector in physical coordinate and $r_{h} $ is halo size. For large halos merging with a single merger such that $m_{2} \gg m_{1} $, $\mu =2$, and $r_{yi}^{} ={r_{h} /2} $ (from Eqs. \eqref{ZEqnNum365738} and \eqref{ZEqnNum581329}). Here $v_{i}^{2} ={GM/8r_{i} } $ (Eq. \eqref{ZEqnNum577609}). Surprisingly, the halo kinetic and potential energies can be derived simply based on the elementary two-body collapse for large halos with an infinitesimal lifetime, where halo density profile information is not even required. 

On the other hand, the halo potential energy can be obtained for halos with a power-law density profile of $\rho _{h} \left(r\right)\sim r^{-m} $, 
\begin{equation} 
\label{ZEqnNum437220} 
P_{h} =-\frac{\int _{0}^{r_{h} }\frac{G}{y}  \left[\int _{0}^{y}\rho _{h} \left(x\right)4\pi x^{2} dx \right]\rho _{h} \left(y\right)4\pi y^{2} dy}{\int _{0}^{r_{h} }\rho _{h} \left(x\right)4\pi x^{2} dx } =-\frac{3-m}{5-2m} \frac{GM}{r_{h} } .
\end{equation} 
The potential energy from an isothermal profile with \textit{m}=2 (Eq. \eqref{ZEqnNum437220}) is exactly consistent with that from the TBCM model in Eq. \eqref{ZEqnNum157977} (the effective potential exponent $n_{e} =-1.5$ for isothermal density \citep[see][Eq. (96)]{Xu:2021-Inverse-mass-cascade-halo-density}. This fact indicates that the isothermal density profile of large halos is a direct result of extremely fast mass accretion or infinitesimal lifetime. In reality, halos have a finite lifetime, and the density profile cannot be exactly isothermal. 

\section{Connections with existing theories}
\label{sec:4}
Solutions developed for TBCM model in Section \ref{sec:3} provide significant insights into existing theories. In this Section, TBCM model is applied to demonstrate the stable clustering hypothesis (SCH). The generalized stable clustering hypothesis (GSCH) is proposed with an interesting scaling for high order moments of pairwise velocity. Finally, connections with violent relaxation and standard spherical collapse model (SCM) are also discussed.

\subsection{Stable clustering hypothesis (SCH) from TBCM and \texorpdfstring{\\}{} generalized SCH for pairwise velocity}
\label{sec:4.1}
The stable clustering hypothesis is a fundamental assumption and one of the few key analytical tools for deeply nonlinear regime of gravitational collapse \citep{Peebles:1980-The-Large-Scale-Structure-of-t}. The dynamic evolution of the density correlation function can be predicted based on this hypothesis and pair conservation equation. The hypothesis states that on sufficiently small scales, there is no stream motion between particles in the physical coordinate. In this case, the peculiar motion cancels out the Hubble flow. The hypothesis equivalently states that the mean pairwise peculiar velocity $\Delta u_{L} $(first order moment) is proportional to the proper separation \textit{r} between pair of particles, i.e $\left\langle \Delta u_{L} \right\rangle =-Hr$. The structure is bound and frozen and the mean particle separation \textit{r} (in physical coordinate) is a constant on sufficiently small scales. In this section, the TBCM model is applied to demonstrate the stable clustering hypothesis and extend it to high order moments of $\Delta u_{L}$. 

The temporal evolution with time scale \textit{s} can be equivalently transformed to the evolution with physical time \textit{t}, where $s=t_{0} \ln \left({t/t_{i} } \right)$ with $t_{i} $ and $t_{0} $ being the initial and current physical time (Eq. \eqref{ZEqnNum185878}). The evolution of comoving size $r_{m} \propto \exp \left(-H_{0} s\right)\propto a^{-1} $ for $n=-1$ in the equilibrium range can be obtained from Eq. \eqref{ZEqnNum758036}, which means a stable clustering frozen in physical coordinate with a fixed proper separation. The stable clustering is only possible for $n=-1$ in expanding background (Eq. \eqref{ZEqnNum758036}). The peculiar pairwise velocity for particle pair with equal mass is defined as \citep[see][Fig. 1]{Xu:2022-The-statistical-theory-of-2nd}
\begin{equation} 
\label{ZEqnNum907578} 
\Delta u_{L} \left(2r\right)=\left(\boldsymbol{\mathrm{u}}_{1} -\boldsymbol{\mathrm{u}}_{2} \right)\cdot \frac{\left(\boldsymbol{\mathrm{x}}_{1} -\boldsymbol{\mathrm{x}}_{2} \right)}{\left|\boldsymbol{\mathrm{x}}_{1} -\boldsymbol{\mathrm{x}}_{2} \right|} ,         
\end{equation} 
which can be directly related to the virial quantity (radial momentum) $G_{s} \left(s\right)$ derived in Eq. \eqref{ZEqnNum463637}. After converting velocity to peculiar velocity with Eq. \eqref{ZEqnNum631619}, the pairwise velocity is (from Eq. \eqref{ZEqnNum479123}),
\begin{equation} 
\label{ZEqnNum390141} 
\begin{split}
a^{{1/2} } \Delta u_{L}&=\Delta v_{L} =2\frac{\boldsymbol{\mathrm{r}}\cdot \boldsymbol{\mathrm{v}}_{1} }{r} =2\frac{G_{s} \left(s\right)}{r} =2\dot{r}\\
&=\frac{2H_{0} r_{i}^{} }{2+n} \exp \left(\frac{-nH_{0} s}{2\left(2+n\right)} \right)\cos \theta _{s} -\frac{2H_{0} r}{2+n},
\end{split}
\end{equation} 
Therefore, for $n=-1$ (using Eq. \eqref{ZEqnNum966949}),
\begin{equation} 
\label{ZEqnNum411840} 
\begin{split}
\Delta u_{L}&=-\frac{2Har}{2+n} +\frac{2H_{0} r_{i}^{} }{2+n} \cos \theta _{s} a_{i}^{\frac{n}{2\left(2+n\right)} } a^{-\frac{n+1}{n+2} }\\ 
&=-2Har+2H_{0} r_{i} a_{i}^{-\frac{1}{2} } \cos \theta _{s}=-2Har+2\beta _{s} u_{i} \cos \theta _{s}.
\end{split}
\end{equation} 

Pairs of particles at time \textit{a} with a separation \textit{r} can be formed at different initial time $a_{i} $ with random initial separations $r_{i} $ and peculiar velocity $u_{i} $. Angles $\theta _{s} $ at a given time \textit{a} can be treated as a random variable (Eq. \eqref{ZEqnNum104758}). Like our treatment of kinetic and potential energies in Eq. \eqref{ZEqnNum187238}, the mean peculiar pairwise velocity for many pairs of particles (ensemble average) is 
\begin{equation} 
\label{ZEqnNum185131} 
\left\langle \Delta u_{L} \right\rangle =-2Har+2\left\langle \beta _{s} u_{i} \cos \theta _{s} \right\rangle .        
\end{equation} 

We may safely assume that $\beta _{s} $, $u_{i} $, and angle $\theta _{s} $ are independent random variables for a sufficiently large number of pairs. The second term on the RHS of Eq. \eqref{ZEqnNum185131} should vanish as $\left\langle \cos \theta _{s} \right\rangle =0$. The first order moment of pairwise velocity is therefore proportional to the separation 2\textit{r} for $r\to 0$ such that 
\begin{equation} 
\label{ZEqnNum128991} 
\left\langle \Delta u_{L} \right\rangle =-2Har=-2a^{-{1/2} } H_{0} r. 
\end{equation} 
Equation \eqref{ZEqnNum128991} is often presented as a direct result of stable clustering hypothesis. There have been many attempts to verify this relation with \textit{N}-body simulations \citep{Efstathiou:1988-Gravitational-Clustering-from-,Colombi:1996-Self-similarity-and-scaling-be}, while here we are able to directly demonstrate this result using the two-body collapse (TBCM) model. 

Similar argument can be extended to higher order moments of pairwise velocity. For second order moment, namely the pairwise velocity dispersion, we have (from Eq. \eqref{ZEqnNum411840})
\begin{equation} 
\label{eq:118} 
\left\langle \Delta u_{L}^{2} \right\rangle \left(r\to 0\right)=4\left\langle \beta _{s}^{2} u_{i}^{2} \cos ^{2} \theta _{s} \right\rangle >0 
\end{equation} 
that is dependent on the exact distributions of $\beta _{s} $, $u_{i} $, and $\theta _{s} $. The non-zero pairwise velocity dispersion is an important signature of the collisionless flow, while $\left\langle \Delta u_{L}^{2} \right\rangle \left(r\to 0\right)=0$ for collisional hydrodynamics where pairs of particles are fully correlated with $r\to 0$ \citep[see][Table 3]{Xu:2022-Two-thirds-law-for-pairwise-ve}. For a uniform distribution of $\theta _{s} $ between [0, 2$\piup$], $\left\langle \cos ^{2} \theta _{s} \right\rangle ={1/2} $. For particle pairs that will form an equilibrium range (stable clustering), a necessary condition is,
\begin{equation} 
\label{eq:119} 
\beta _{s} =\frac{H_{0} r_{i} }{v_{i} } \le \beta _{s1} =1 
\end{equation} 
\begin{equation} 
\label{eq:120} 
\left\langle \Delta u_{L}^{2} \right\rangle \left(r\to 0\right)=2H_{0}^{2} \left\langle r_{i}^{2} a_{i}^{-1} \right\rangle =2\left\langle \beta _{s}^{2} u_{i}^{2} \right\rangle .       
\end{equation} 

The higher order moments of pairwise velocity with $r\to 0$ can be similarly derived from Eq. \eqref{ZEqnNum411840}, where the even and odd moments can be obtained up to the first order of \textit{r},
\begin{equation} 
\label{ZEqnNum346814} 
\left\langle \Delta u_{L}^{2m} \right\rangle \left(r\to 0\right)=\left(2H_{0} \right)^{2m} \left\langle r_{i}^{2m} \right\rangle \left\langle a_{i}^{-m} \right\rangle \left\langle \cos ^{2m} \theta _{s} \right\rangle ,      
\end{equation} 
\begin{equation} 
\label{ZEqnNum449713} 
\left\langle \Delta u_{L}^{2m+1} \right\rangle \left(r\to 0\right)=-2Har\left(2H_{0} \right)^{2m} \left\langle r_{i}^{2m} \right\rangle \left\langle a_{i}^{-m} \right\rangle \left\langle \cos ^{2m} \theta _{s} \right\rangle .      
\end{equation} 
A simplified relation is found between the odd and even pairwise velocity moments with $r\to 0$, 
\begin{equation} 
\label{ZEqnNum921059} 
\left\langle \Delta u_{L}^{2m+1} \right\rangle =\left(2m+1\right)\left\langle \Delta u_{L}^{2m} \right\rangle \left\langle \Delta u_{L}^{} \right\rangle =\left(2m+1\right)\left\langle \Delta u_{L}^{2m} \right\rangle \left(-2Har\right),    
\end{equation} 
which reduces to the standard stable clustering hypothesis (Eq. \eqref{ZEqnNum128991}) for $m=0$. Equation \eqref{ZEqnNum921059} can be considered as a generalized stable clustering hypothesis for pairwise velocity that can be directly confirmed by N-body simulations.

\begin{figure}
\includegraphics*[width=\columnwidth]{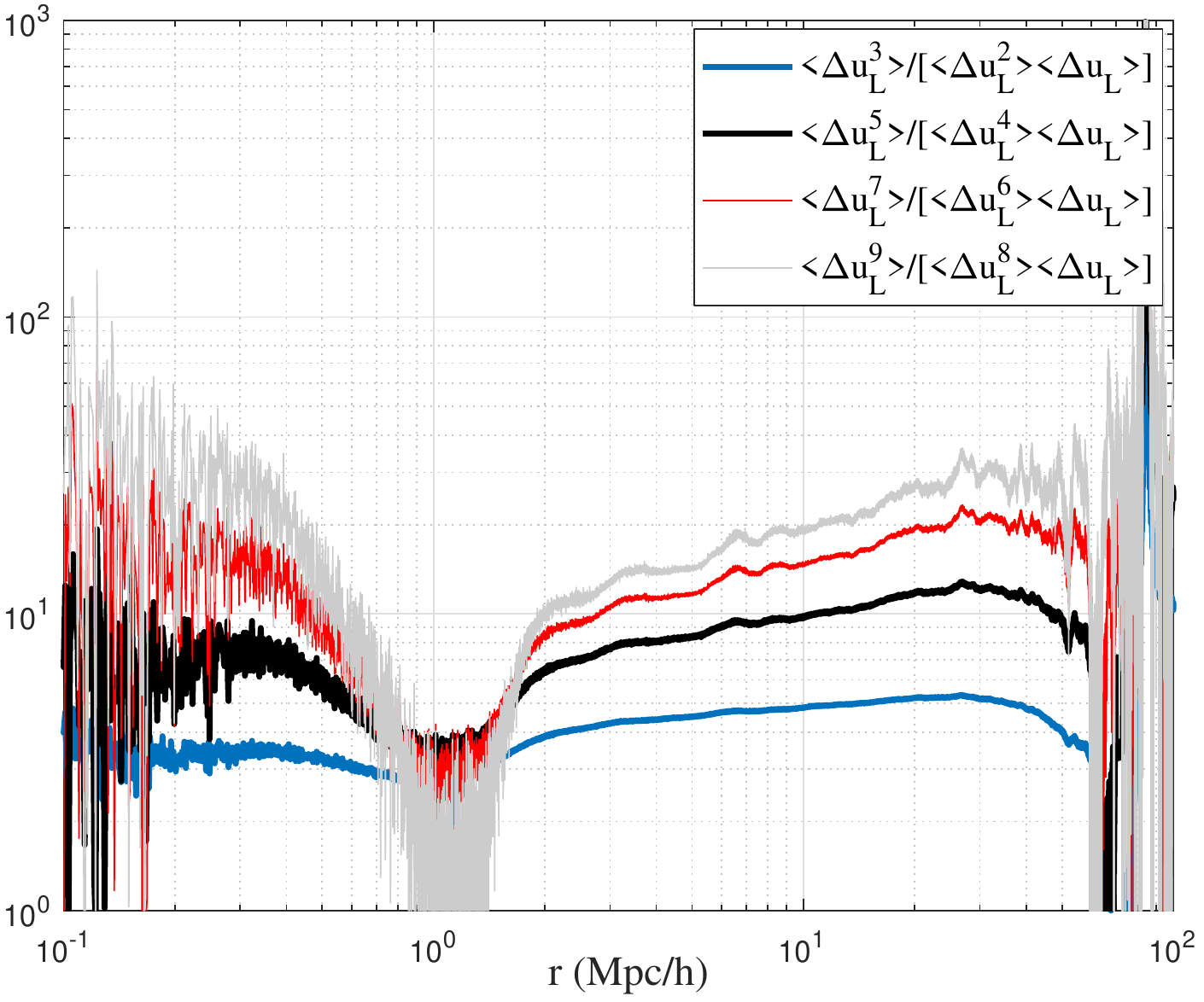}
\caption{The ratio between (2m+1)th odd and (2m)th even order moments of pairwise velocity from N-body simulation. This ratio is predicted by generalized stable clustering hypothesis (GSCH in Eq. \eqref{ZEqnNum921059}) to be (2m+1). Comparison between prediction of GSCH and N-body simulation is presented in Fig. \ref{fig:8} for high order moments of pairwise velocity.}
\label{fig:7}
\end{figure}

\begin{figure}
\includegraphics*[width=\columnwidth]{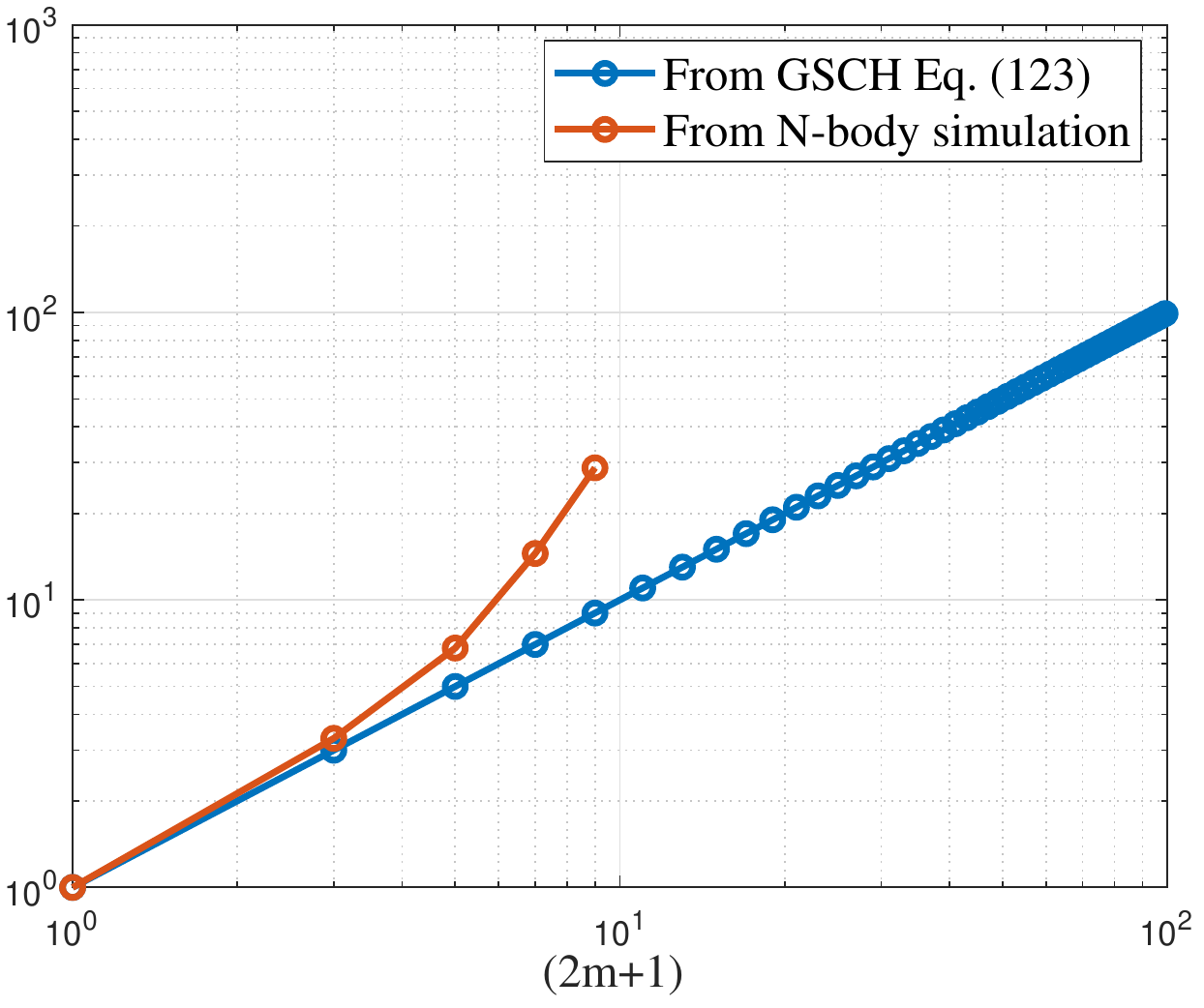}
\caption{The comparison of predicted ratio from generalized stable clustering hypothesis in Eq. \eqref{ZEqnNum921059} and those from N-body simulation (Fig. \ref{fig:7}) for (2\textit{m}+1)th moments of pairwise velocity. The deviation might require further study using N-body simulations with higher resolution to sample the statistics of particle pairs on small scale.}
\label{fig:8}
\end{figure}

In N-body simulation, all particle pairs with a given separation $r$ were identified. The moments of pairwise velocity is computed as the average for all pairs of particles with the same $r$ (\citep[also see][Fig. 24]{Xu:2022-Two-thirds-law-for-pairwise-ve}). Figure \ref{fig:7} presents the plot of Eq. \eqref{ZEqnNum921059} from N-body simulation, i.e. the ratio between odd and even moments of pairwise velocity $\langle\Delta u_{L}^{2m+1}\rangle/(\langle\Delta u_{L}^{2m}\rangle\langle\Delta u_L\rangle)$. Figure \ref{fig:8} presents the comparison of that ratio with predicted value of 2\textit{m}+1. The deviation from prediction for higher order moments might come from the spatial intermittence of energy cascade and require N-body simulations with higher resolution to reduce the large fluctuation at small $r$ in Fig. \ref{fig:7}. The spatial intermittence of energy cascade in dark matter flow leads to different rate of energy cascade for different halos and affects the small scale dynamics \citep[see][Fig. 9]{Xu:2022-The-baryonic-to-halo-mass-rela}.

The 2\textit{m}th order generalized kurtosis of the PDF (Probability Distribution Function) of pairwise velocity $\Delta u_{L} $ is defined as, 
\begin{equation} 
\label{eq:124} 
K_{2m} ={\left\langle \Delta u_{L}^{2m} \right\rangle /\left\langle \Delta u_{L}^{2} \right\rangle ^{m} } .         
\end{equation} 
Let's assume the second order moment has a general form of $\left\langle \Delta u_{L}^{2} \right\rangle =\alpha _{u} u_{0}^{2} a^{\beta _{u} } $ when $r\to 0$, where $u_{0}^{2} $ is the one-dimensional velocity dispersion of the entire N-body system at present epoch. The high order moments and generalized kurtosis of the pairwise velocity for $r\to 0$ can be obtained as (from Eq. \eqref{ZEqnNum921059}),
\begin{equation} 
\label{eq:125} 
\left\langle \Delta u_{L}^{2m} \right\rangle =K_{2m} \alpha _{u}^{m} u_{0}^{2m} a^{\beta _{u} m} ,          
\end{equation} 
\begin{equation} 
\label{eq:126} 
\left\langle \Delta u_{L}^{2m+1} \right\rangle =-2\left(2m+1\right)K_{2m} \alpha _{u}^{m} u_{0}^{2m+1} a^{{\beta _{u} m-\frac{1}{2} }} \frac{H_{0} r}{u_{0}}<0,     
\end{equation} 
\begin{equation} 
\label{eq:127} 
K_{2m+1} \left(r\right)=-2\left(2m+1\right)K_{2m} \alpha _{u}^{-\frac{1}{2} } a^{{-\frac{\left(1+\beta _{u} \right)}{2}} }  \frac{H_{0} r}{u_{0}}<0,     
\end{equation} 
where $\alpha _{u} $, $\beta _{u} $, and $K_{2m} $ fully determine all these moments. Simulations suggest $\alpha _{u} =2$ and $\beta _{u} ={3/2} $. 

The limiting velocity of dark matter particles follows a $X$ distribution with a Gaussian core and exponential wing to maximize system entropy \citep[see][Fig. 4]{Xu:2021-The-maximum-entropy-distributi}, while the probability distribution of pairwise velocity $\Delta u_{L}$ with $r\to 0$ can be different with kurtosis $K_{2m}$ analytically derived \citep[see][Section 5.2]{Xu:2022-Two-thirds-law-for-pairwise-ve}. The generalized stable clustering hypothesis (GSCH) from TBCM model shows that on small scale with $r\to 0$, the odd moments of $\Delta u_{L} $ are always proportional to \textit{r} while the even order moments are independent of the separation \textit{r}. Especially, the second moment of pairwise velocity (pairwise velocity dispersion) follows a two-thirds law $\langle \Delta u_{L}^2 \rangle \propto r^{2/3}$ on small scale \citep{Xu:2022-Two-thirds-law-for-pairwise-ve} that might be used to derive dark matter particle mass and properties \citep{Xu:2022-Postulating-dark-matter-partic}.

\subsection{Connections with violent relaxation}
\label{sec:4.2}
The violent relaxation is originally proposed for the collisionless system with a time-dependent potential to explain the absence of tendency to segregate different masses during the relaxation \citep{Lyndenbell:1967-Statistical-Mechanics-of-Viole}. The TBCM model can be considered as a special example of violent relaxation involving only two masses. The evolution of mean separation $r_{m} \left(s\right)$ (Eq. \eqref{ZEqnNum758036}) do not involve particle mass. Particle mass only affects the frequency term $\omega _{m} \left(s\right)$ through initial velocity $v_{i} $ (Eq. \eqref{ZEqnNum870001}). The characteristic time of relaxation (in scale of \textit{s}) is only dependent on $H_{0} $, regardless of particle mass (see energy evolution in Eqs. \eqref{ZEqnNum521799} to \eqref{ZEqnNum187238}). Two particles with unequal masses collapse at the same rate with the same characteristic time of relaxation such that this type of relaxation does not lead to mass segregation. We may examine the energy transfer between two particles during a two-body collapse. The initial ratio of kinetic and potential energy between two particles are (from Eqs. \eqref{ZEqnNum860465} and \eqref{ZEqnNum194669})
\begin{equation} 
\label{ZEqnNum132983} 
\frac{K_{si1} }{K_{si2} } =\frac{m_{1} v_{1i}^{2} }{m_{2} v_{2i}^{2} } =\frac{m_{1} \mu ^{2} }{m_{2} \left(2-\mu \right)^{2} } =\frac{m_{2} }{m_{1} } ,  \frac{P_{si1} }{P_{si2} } =1. 
\end{equation} 
The kinetic energy of two particles evolves during the two-body collapse is (from Eq. \eqref{ZEqnNum369546}),
\begin{equation} 
\label{eq:129} 
K_{s1} =\frac{1}{2} m_{1} \left(v_{x1}^{2} +v_{y1}^{2} \right)=\frac{1}{2} m_{1} \mu ^{2} \left[\dot{r}^{2} +r^{2} F\left(s\right)^{-4} \right],      
\end{equation} 
and
\begin{equation} 
\label{eq:130} 
K_{s2} =\frac{1}{2} m_{2} \left(v_{x2}^{2} +v_{y2}^{2} \right)=\frac{1}{2} m_{2} \left(2-\mu \right)^{2} \left[\dot{r}^{2} +r^{2} F\left(s\right)^{-4} \right].     
\end{equation} 

Obviously, the ratio of kinetic energy between two particles is time-invariant and equals the initial ratio in Eq. \eqref{ZEqnNum132983}. The potential energy of two particles also evolves with a constant ratio of 1. There is no energy transfer between two particles with unequal mass. Therefore, the energy equipartition does not apply here for particles with different masses. For comparison, during a collisional relaxation, the relaxation time of massive particles is less than that of light particles (inversely proportional to particle mass for a two-body relaxation process \citep{Leigh:2013-Modifying-two-body-relaxation-}. The energy equipartition enables transferring of kinetic energy between different particles such that massive and light particles share the same kinetic energy. Therefore massive particles have small velocity and tend to fall to the center of structure \citep{Spitzer:1969-Equipartition-and-the-Formatio}.

\subsection{Connections with spherical collapse model (SCM)}
\label{sec:4.3}
The spherical collapse model (SCM) solves the motion of spherical shells of matter surrounding an over-density, where many important insights can be obtained for highly-nonlinear gravitationally collapsing objects \citep{Gunn:1972-Infall-of-Matter-into-Clusters}. This section will reveal some fundamental connections between TBCM and SCM. The equation of motion for a SCM model in physical coordinates reads
\begin{equation} 
\label{ZEqnNum587263} 
\frac{d^{2} R}{dt^{2} } =-\frac{GM}{R^{2} } ,          
\end{equation} 
where \textit{R} is the radius of the spherical shell and $M$ is the mass enclosed by that shell. The initial velocity of mass shell is assumed to be Hubble flow,
\begin{equation} 
\label{eq:132} 
\left. \frac{dR}{dt} \right|_{t=0} =H_{i} R_{i} =H_{0} x_{i} a_{i}^{-{1/2} }  
\end{equation} 
where $x_{i} =x\left(t=0\right)$ is the initial radius in a comoving system. Solution to Eq. \eqref{ZEqnNum587263} can be written in a parametric form,
\begin{equation}
R=A\left(1-\cos \theta \right) \quad \textrm{and} \quad t=B\left(\theta -\sin \theta \right),      
\label{eq:133}
\end{equation}

\noindent where two constants \textit{A} and \textit{B} are related to the initial radius $x_{i} $ in comoving coordinates, 
\begin{equation}
x_{i} =x\left(t=0\right)={A\left(12\pi \right)^{{2/3} } / 2} \quad \textrm{and} \quad A^{3} =GMB^{2}. 
\label{ZEqnNum468309}
\end{equation}

For a direct comparison, the SCM model (Eq. \eqref{ZEqnNum587263}) can be equivalently expressed in the transformed system with comoving coordinate \textit{x} and transformed time scale \textit{s},
\begin{equation} 
\label{ZEqnNum965142} 
\frac{\partial ^{2} x}{\partial s^{2} } +\frac{H_{0} }{2} \frac{\partial x}{\partial s} +\frac{GM}{x^{2} } =\frac{H_{0}^{2} }{2} x.         
\end{equation} 

By setting $x=2r$ (SCM models a spherical over-density with uniform mass distribution), the SCM Eq. \eqref{ZEqnNum965142} can be rewritten as
\begin{equation} 
\label{ZEqnNum897023} 
\frac{\partial ^{2} r}{\partial s^{2} } +\frac{H_{0} }{2} \frac{\partial r}{\partial s} +\frac{GM}{2\left(2r\right)^{2} } =\underbrace{\frac{H_{0}^{2} }{2} r}_{1}.        
\end{equation} 

The TBCM model presented in Section \ref{sec:3.1} (Eq. \eqref{ZEqnNum332742}) describes a two-body system in an expanding background with uniform background density. The equation reads
\begin{equation} 
\label{ZEqnNum814748} 
\frac{\partial ^{2} r}{\partial s^{2} } +\frac{H_{0} }{2} \frac{\partial r}{\partial s} +\frac{GM}{2\left(2r\right)^{2} } =\underbrace{\frac{\left(r_{i} v_{i} \right)^{2} }{r^{3} } \exp \left(-H_{0} s\right)}_{2},       
\end{equation} 
where $M=m_{1} +m_{2} $ is the total mass of two-body system. The SCM model essentially describes a self-gravitating system in an otherwise empty universe (Eq. \eqref{ZEqnNum587263}). The energy is conserved for SCM model in physical coordinates. By comparing the SCM Eq. \eqref{ZEqnNum897023} with the TBCM Eq. \eqref{ZEqnNum814748}, the original SCM model has an extra force term on the RHS (term 1 in Eq. \eqref{ZEqnNum897023}) that is due to the absence of a uniform background density. The TBCM model has a time-varying frequency force due to the angular momentum (term 2 in Eq. \eqref{ZEqnNum814748}). 

The original SCM can be considered to describe exactly a two-body collapse in an otherwise empty universe, with one-dimensional radial motion only and zero angular momentum. The initial conditions for original SCM model (Eq. \eqref{ZEqnNum897023}) is an initial separation $r_{i} ={x_{i} /2} $ (Eq. \eqref{ZEqnNum468309}) and a zero initial velocity in a transformed system. The modified non-radial SCM model introduces an additional constant centrifugal force to indirectly account for the effect of non-radial motion \cite{White:1992-Models-for-Galaxy-Halos-in-an-,Nusser:2001-Self-similar-spherical-collaps}, while it still models a self-gravitating system in an otherwise empty universe.   

Equivalently, the proposed TBCM model in expanding background can be considered as a spherical non-radial collapse model describing the gravitational collapse of a mass shell with a non-zero angular momentum and non-radial orbits (non-radial spherical collapse model). Both models predict a critical halo density ratio of $\Delta_c=18\pi ^{2} $ to the background density. However, the original SCM model cannot predict the existence of an equilibrium range (stable clustering for $n=-1$). In SCM model, the system is out of virial equilibrium initially and reaching the virial equilibrium at the critical density, where effect of halo mass accretion cannot be explicitly considered.  

For comparison, the TBCM model allows the existence of an equilibrium range for $\beta _{s} \le \beta _{s1} $,  where the initial density is at least twice the background density (Eqs. \eqref{ZEqnNum552089} and \eqref{ZEqnNum776754}). The TBCM model can be considered as the elementary step for mass accretion/cascade. Halos with an infinitesimal lifetime (due to the fast mass accretion) approach a critical density with a ratio of $\Delta_c=18\pi ^{2}$ to the background, where $\beta _{s} $ converges to the critical value $\beta _{s2} ={1/3\pi}$ (Eqs. \eqref{ZEqnNum937360} and \eqref{ZEqnNum776754}). The stable clustering hypothesis (SCH) can be demonstrated by the TBCM model and generalized to high order moments of pairwise velocity. Richer information on halo energy/momentum/structure can be obtained from a TBCM model.

\section{Conclusions}
\label{sec:5}
A transformed system for equation of motion is proposed by introducing a different time scale \textit{s} for the motion of collisionless particles in expanding background (Eq. \eqref{ZEqnNum730753}). The equivalence with the original comoving systems is established. A two-body collapse model (TBCM) for gravity with an arbitrary exponent \textit{n} is formulated in the transformed system (Eq. \eqref{ZEqnNum365738}). Results obtained can be readily translated back to the original system. A complete analysis of TBCM model is provided with governing equations for radius function (Eq. \eqref{ZEqnNum332742}) or frequency function (Eq. \eqref{ZEqnNum502680}). The original five model parameters, i.e. the potential exponent \textit{n}, Hubble constant $H_{0} $, initial size $r_{i} $ and velocity $v_{i} $, and system mass $M=m_{1} +m_{2} $, can be grouped into three dimensionless parameters \textit{n}, $\beta _{s} ={H_{0} r_{i} /v_{i} } $, and $\gamma _{s} =\left({v_{ri} /v_{i} } \right)^{2} $ (Eqs. \eqref{ZEqnNum129548}, \eqref{ZEqnNum629408}). Here $v_{ri} $ is the circling velocity for a given $r_{i} $ and $M$, and\textit{ n} in static background without damping. The competition between gravitational force, expanding background (damping), and angular momentum classifies the two-body collapse into two regimes (Fig. \ref{fig:5}): 1) a free fall collapse without oscillatory motion for weak angular momentum; and 2) an equilibrium collapse with oscillations for weak damping, when $\beta _{s} \ll 1$, $-2<n<0$ and $\gamma _{s} \sim 1$. Two regimes are studied as follows.

For free fall collapse, the free fall time $t_{c} $ in an expanding background can be analytically derived as a function of the free fall time $s_{ce} $ in static background and the beginning time $t_{i} $ of free fall (Fig. \ref{fig:6} and Eq. \eqref{ZEqnNum190681}). The two-body collapse can have a greater free fall time $t_{c} $ if two-body system begins to collapse at an earlier time $t_{i}$ due to larger Hubble parameter (or damping) $H$. 

For an equilibrium collapse, solutions identify three distinct regimes (transitional, equilibrium, and final collapse in Fig. \ref{fig:5}). An exponential evolution of two-body system size, energy and momentum can be obtained in transformed system (Fig. \ref{fig:4}, Eqs. \eqref{ZEqnNum248861}, \eqref{ZEqnNum521799}, \eqref{ZEqnNum978254}, \eqref{ZEqnNum387242}, \eqref{ZEqnNum463637} and \eqref{ZEqnNum129930}), or equivalently a power-law evolution in the original  comoving system. A critical value of $\beta _{s} \le \beta _{s1} $ (Eq. \eqref{ZEqnNum552089}) is required for the existence of an equilibrium range. Equivalently, a maximum system size or a minimum initial velocity can be identified (Eq. \eqref{ZEqnNum562446}). 

The second critical value of $\beta _{s} =\beta _{s2} $ (Eq. \eqref{ZEqnNum937360}) can be identified for large halos with an infinitesimal lifetime (due to fast mass accretion). Large halos tend to be synchronized and generated at the same time with small dispersion in their properties. Conversely, small halos tend to have longer lifetime with more diversified properties at a given redshift. The two-body angular velocity (Eq. \eqref{ZEqnNum762297}), typical orbital period (Eq. \eqref{ZEqnNum256922}), the angle of incidence (ratio between radial and circular velocity) (Eq. \eqref{ZEqnNum994889}), and the critical density ratio of 18$\pi^2$ (Eq. \eqref{ZEqnNum979432}) can all be derived from the TBCM model. Isothermal density profile is also a direct result of an infinitesimal lifetime.

Finally, the TBCM model demonstrates the stable clustering hypothesis (SCH) for an equilibrium collapse, where mean pairwise velocity is proportional to separation (Eqs. \eqref{ZEqnNum390141} and \eqref{ZEqnNum128991}). A generalized stable clustering hypothesis (GSCH) is also developed for higher odd and even order moments that are related by mean pairwise velocity (Eqs. \eqref{ZEqnNum346814}, \eqref{ZEqnNum449713}, \eqref{ZEqnNum921059} and Figs. \ref{fig:7} and \ref{fig:8}). The two-body collapse in expanding background is independent of particle masses, where the energy equipartition does not apply. Compared to the original spherical collapse model (SCM), the TBCM model can be naturally considered as a spherical non-radial collapse model with non-zero angular momentum (Eqs. \eqref{ZEqnNum897023} and \eqref{ZEqnNum814748}). Both models predict the same critical halo density ratio of 18$\pi^2$, while the original SCM model cannot predict a stable clustering. The TBCM model also suggests a power-law energy evolution on large scale that will be further investigated \citep{Xu:2022-The-evolution-of-energy--momen}. 


\section*{Data Availability}
Two datasets underlying this article, i.e. a halo-based and correlation-based statistics of dark matter flow, are available on Zenodo \citep{Xu:2022-Dark_matter-flow-dataset-part1,Xu:2022-Dark_matter-flow-dataset-part2}, along with the accompanying presentation slides "A comparative study of dark matter flow \& hydrodynamic turbulence and its applications" \citep{Xu:2022-Dark_matter-flow-and-hydrodynamic-turbulence-presentation}. All data files are also available on GitHub \citep{Xu:Dark_matter_flow_dataset_2022_all_files}.

\bibliographystyle{mnras}
\bibliography{Papers}
\label{lastpage}
\end{document}